\documentclass[preprint,showpacs,prd,amsmath,amssymb]{revtex4}
\usepackage{graphicx}
\usepackage{dcolumn}

\renewcommand{\bar}[1]{\overline{#1}}

\begin{document}

\preprint{USM-TH-118} 

\title{ Evidence for SU(3) symmetry breaking from hyperon production}

\author{Jian-Jun Yang}
\email{jjyang@fis.utfsm.cl}
\affiliation{Department of Physics, Nanjing Normal
University, Nanjing 210097, China\\
Departamento de F\'\i sica, Universidad T\'ecnica Federico 
Santa Mar\'\i a, Casilla 110-V, Valpara\'\i so, Chile
\footnote{Maillng address}}

\begin{abstract}

We examine the SU(3) symmetry breaking in hyperon semileptonic
decays (HSD) by considering two typical sets of quark
contributions to the spin content of the octet baryons:
Set-1 with SU(3) flavor symmetry and Set-2 with SU(3) flavor
symmetry breaking in HSD. The quark distributions of the octet baryons
are calculated with a successful statistical model.
Using an approximate relation between the quark fragmentation 
functions and the quark distributions, 
we predict polarizations of the octet baryons 
produced in $e^+e^-$ annihilation and semi-inclusive deeply 
lepton-nucleon scattering in order to reveal  
the SU(3) symmetry breaking effect on the spin structure 
of the octet baryons. We find that the SU(3) symmetry breaking 
significantly affects the hyperon polarization. The available 
experimental data on the $\Lambda$ polarization seem to favor 
the theoretical predictions with SU(3) symmetry breaking. 
We conclude that there is a possibility to get a collateral 
evidence for SU(3) symmetry breaking from hyperon production.
The theoretical errors for our predictions are discussed.

\end{abstract}


\maketitle

\section{Introduction}

The proton spin has stayed as an interesting problem for more than
one decade. The quark-spin content of the nucleon is usually
extracted according to the data on polarized deep inelastic
lepton-nucleon scattering and hyperon semileptonic decays (HSD)
with the SU(3) flavor symmetry assumption.
However, some current analyses have shown
that the effect of SU(3) symmetry breaking in HSD  on the
quark-spin content of the octet baryons is
significant~\cite{KimPRD, Manohar98, KimNPA}.
The effect has been estimated  by the chiral quark
soliton model~\cite{KimPRD} and the large
$N_c$ QCD~\cite{Manohar98}. The
consistent results were obtained  separately by different approaches.
However, there is a lack of external evidences for 
the SU(3) symmetry breaking. In order to view the SU(3) flavor symmetry 
breaking effect as a whole, we should broaden our vision from 
the nucleon to all octet baryons  since  the effect is related to
the  octet hyperon semileptonic decays. Hyperon production 
is a very important source from which we can get information on
the spin properties of the octet hyperons.     
The $\Lambda$ hyperon is of special interest in this respect since its
decay is self-analyzing with respect to its spin direction
due to the dominant weak decay $\Lambda
\to p \pi^-$ and the particularly large asymmetry of the angular
distribution of the decay proton in the $\Lambda$ rest frame. So
polarization measurements are relatively simple to be performed
and the polarized fragmentation functions of quarks to the $\Lambda$
can be measured. Also the fragmentation of quarks to  $\Sigma$ and $\Xi$
hyperons can be investigated experimentally
since the detection technique of $\Sigma$ and $\Xi$ hyperons is
gradually  maturing and will allow
to measure various quark to hyperon fragmentation 
functions~\cite{SigmaP,XiP,Sigma0P}.
On the other hand, recent investigations indicate that 
the quark fragmentation functions seem to have relation to the corresponding
quark distributions via an approximate relation~\cite{GLR}.
Thus, it is of great significance  to examine  the influence
of the SU(3) symmetry breaking on the spin properties
of the octet hyperons from fragmentation.

 Based on the known  experimental data of HSD
constants and the first moment of the spin structure function of the
proton in DIS, we can extract quark contributions to the spin content of
the octet baryons. In order to obtain quark distributions of the 
octet baryons, we need some theoretical models. 
Quark distributions of the nucleon have  been
precisely measured by combination of various deep inelastic scattering (DIS)
processes~\cite{DISprocess} and Drell-Yan processes~\cite{DYprocess}.
With decades of experiments, our knowledge of the quark distributions for the nucleon
is more or less clear concerning
the bulk features of momentum, flavor and helicity distributions. 
For the other octet baryons,
we can only predict the quark distributions for them by means of  some successful models which produce
the quark structure of the nucleon.
Recently, a statistical model for polarized and unpolarized parton distributions of the
nucleon has been  presented by Bhalerao {\it{ et
al.}}~\cite{Bhalerao00}. The model can reproduce the almost all data on the nucleon
structure functions $F_2^p(x,Q^2)$,
$F_2^p(x)-F_2^n(x)$ and parton sum rules.  We find that the model
 is suitable to study the SU(3) symmetry breaking effect  
 on the  quark spin properties
 of the octet baryons since quark contributions to 
 the spin content are  important inputs of the model.
  According to some constraints,
we determine two typical sets of  parton  density  functions (PDFs) for the
octet baryons  at an initial scale: Set-1 with SU(3)  flavor symmetry and
Set-2 with SU(3) flavor symmetry breaking in HSD. Then we relate PDFs to
fragmentation functions at the initial scale by using  an approximate  relation between the quark
distributions and  the fragmentation functions.
 Finally, we employ the evolved
two sets of fragmentation functions to predict octet 
baryon polarizations in $e^+e^-$ annihilation, the polarized 
charged lepton DIS and neutrino DIS in order to reveal the 
influence of SU(3) symmetry breaking on baryon production.

The  paper  is arranged  as follows: In Sec.~II,  we will re-extract
quark contributions to the spin content of the octet baryons  based on some
known experimental data in order to obtain reasonable central values 
for them. In Sec.~III we present the quark spin structure  of all octet baryons
based on a successful statistical model at an initial scale.
In Sec.~IV, we relate the fragmentation functions to the corresponding
quark distributions and evolve them to a higher energy scale.
Then, we predict spin observables for the octet baryons
produced in $e^+e^-$ annihilation, charged letopn DIS and neutrino DIS.
We find that one can get a possible collateral evidence for SU(3) symmetry 
breaking from hyperon production.
A brief summary with some discussions on the theoretical uncertainties
is given in Sec.~V.

\section{Quark Contributions to Spin Content of  Octet Baryons}

The standard way to determine quark contributions to the spin content of a baryon in the
$J^P= \frac{1}{2}$ octet is based on  two pieces of information.
One comes from the first moment of the spin structure function
$g_1^p(x)$ of the proton:
\begin{equation}
I_p=\int \limits_0^1 g_1^p (x) d x = \frac{1}{18} (4 \Delta U
+\Delta D + \Delta S) (1-\frac{\alpha_s}{\pi}+\cdots),
\end{equation}
which is obtained in deep-inelastic lepton-proton scattering
experiments. According to the experimental data of
$I_p$~\cite{EMC89}, one has

\begin{equation}
\Gamma_p = 4 \Delta U + \Delta D + \Delta S = 2.42 \pm 0.26.
\end{equation}
Another information  comes from the hyperon semileptonic
decay constants $F$ and $D$ which are usually obtained  from
the empirical values for the ratios of axial-vector to vector
coupling constants $g_A/g_V$  via

\begin{equation}
(g_A/g_V)^{(n \to p)} = F + D, \hspace{0.5cm}
(g_A/g_V)^{(\Sigma^- \to n)} = F - D
\end{equation}
with the exact flavor SU(3) symmetry. However, there are some uncertainties
in the analysis since $F$ and $D$ can also be obtained
by combining any another two ratios of axial-vector to vector
coupling constants $g_A/g_V$ of six known weak semileptonic
decays~\cite{HSD1,HSD2}. With various combinations, it is
found that there exist large
uncertainties  in the central values of $F$ and $D$ as follows:

\begin{equation}
F=0.40 \div 0.55, \hspace{0.5cm}
D=0.70 \div 0.89,
\end{equation}
which shows that {\it the theoretical error due to using the exact SU(3)
symmetry in  describing  the hyperon semileptonic decays is about 15\%}.
Thus, the obtained the quark spin content of the proton
$\Delta \Sigma =\Delta U + \Delta D + \Delta S$
based on the constants $F$ and $D$ can be any value in the range
$[0.02, 0.30]$, which implies that the SU(3)
symmetry breaking plays an essential role in
extracting $\Delta \Sigma$ from the experimental data.
For this reason,  the SU(3) symmetry breaking effect has been recently
considered in the chiral quark soliton model and the large $N_c$ QCD.
The same algebraical structure of ratios of axial-vector to
vector coupling constants for known weak semileptonic decays
with dynamical parameters were
obtained by two different approaches~\cite{KimPRD, Manohar98}.
In order to obtain model independent results,
one  can fix  the dynamical parameters by fitting  them
to  the data of the known six
weak semileptonic decays  instead  of calculating them
within a specified model.
The quark contributions $\Delta Q$ ($Q=U$, $D$, $S$) to the spin
content of the octet baryons given by Kim {\it et al}~\cite{KimNPA} in
this way are listed in Table~\ref{table1}.

\begin{table}
\caption{ $\Delta Q$ in the SU(3) broken analysis of Ref.~\cite{KimNPA}.}
\begin{tabular}{|c||c|c|c|}\hline
Baryon &  $~~~\Delta U~~~$ & $~~~\Delta  D~~~$ & $~~~\Delta S~~~$ \\ \hline
 ~~~~p~~~~ &
 ~~ 0.72 $\pm$ 0.07~~ & ~~-0.54 $\pm$ 0.07~~ & ~~ 0.33 $\pm$ 0.51~~ \\ \hline
 ~~~~n~~~~ &
 ~~-0.54 $\pm$ 0.07~~ & ~~ 0.72 $\pm$ 0.07~~ & ~~ 0.33 $\pm$ 0.51~~ \\ \hline
 ~~~~$\Sigma^+$~~~~ &
 ~~ 0.73 $\pm$ 0.17~~ & ~~-0.37 $\pm$ 0.19~~ & ~~-0.18 $\pm$ 0.39~~ \\ \hline
 ~~~~$\Sigma^0$~~~~ &
 ~~ 0.18 $\pm$ 0.08~~ & ~~ 0.18 $\pm$ 0.08~~ & ~~-0.18 $\pm$ 0.39~~ \\ \hline
 ~~~~$\Sigma^-$~~~~ &
 ~~-0.37 $\pm$ 0.19~~ & ~~ 0.73 $\pm$ 0.17~~ & ~~-0.18 $\pm$ 0.39~~ \\ \hline
 ~~~$\Lambda^0$~~~~ &
 ~~-0.02 $\pm$ 0.17~~ & ~~-0.02 $\pm$ 0.17~~ & ~~ 1.21 $\pm$ 0.54~~ \\ \hline
  ~~~~$\Xi^-$~~~~ &
 ~~ 0.02 $\pm$ 0.16~~ & ~~-0.14 $\pm$ 0.21~~ & ~~ 1.50 $\pm$ 0.60~~ \\ \hline
 ~~~$\Xi^0$~~~~ &
 ~~-0.14 $\pm$ 0.21~~ & ~~ 0.02 $\pm$ 0.16~~ & ~~ 1.50 $\pm$ 0.60~~ \\ \hline
\end{tabular}\label{table1}
\end{table}

One can see  from Table~\ref{table1} that the results have very big errors
and  some central values of them are beyond the physics region
although they could be meaningful with their errors. The central
values of the total quark spin content of the  $\Lambda$ and $\Xi$ hyperons
are larger than 1.
In addition, the quark spin content $\Delta \Sigma = 0.51 $
of the nucleon is out of the range $[0.02, 0.3]$ which corresponds
to different combinations of the known 6 semileptonic decay constants.
The large errors in the results are mainly due to a very 
large experimental error in the data on the  $\Xi^- \to \Sigma^0$ decay. 
In order to improve the central values of quark 
contributions to the spin content, we re-do the  
analysis by adopting  the data on the
first moment $I_p$ of the proton spin structure function $g_1^p$
instead of using  the data on the  $\Xi^- \to \Sigma^0$ decay since our
knowledge on $g_1^p$ seems to be a little better than 
on $(g_A/g_V)^{(\Xi^- \to \Sigma^0 )}$.
Following Ref.~\cite{KimNPA}, we can express ratios
of axial-vector to vector coupling constants for  5
semileptonic decays and $\Gamma_p$ as:

\begin{equation}
A_1=(g_A/g_V)^{(n \to p)} = -14 r +2 s-44 x-20 y-4 z +8q,
\end{equation}

\begin{equation}
A_2=(g_A/g_V)^{(\Sigma^+ \to \Lambda )} = -9 r -3 s -42 x -6 y +15 q,
\end{equation}

\begin{equation}
A_3=(g_A/g_V)^{(\Lambda \to p )} = -8 r +4 s +24 x -2 z -6 q,
\end{equation}

\begin{equation}
A_4=(g_A/g_V)^{(\Sigma^- \to n )} = 4 r+ 8 s -4 x -4 y +2 z + 4 q,
\end{equation}

\begin{equation}
A_5=(g_A/g_V)^{(\Xi^- \to \Lambda )} = -2 r+6 s -6 x +6 y -2 z + 6 q,
\end{equation}

\begin{equation}
\Gamma_p = -24 r +132 s -48 x -66 y +6 z +48 q,
\end{equation}
where $r$, $s$, $x$, $y$, $z$ and $q$ are dynamical parameters
within the chiral quark model. We can fix these parameters
by solving the above set of  equations with the central
values of the experimental data~\cite{EMC89,HSD1,HSD2}.
The six parameters are found to be: $x=0.00035405$,
$s=0.00398698$, $y=0.00006039$, $q=0.00214004$, $z=-0.02560072$, 
$r=-0.08189920$. With the obtained parameters, we find

\begin{equation} (g_A/g_V)^{(\Xi^- \to \Sigma^0 )} = -14 r+2 s
+22 x +10 y +2 z -4 q = 1.103,
\end{equation}
which is close to the experimental data $1.278 \pm 0.158$.
With our new set of parameters, quark contributions to the spin
content of the octet baryons as listed  in Table~\ref{table3} are in the range
of physics region. In addition, the  total spin contribution
$\Delta \Sigma = 0.08 $ for the nucleon is
in  the interval  $[0.02, 0.3]$  which covers the range obtained by
different combinations of the known 6 semileptonic decay constants.
All of these indicate that the above obtained central values of 
quark contributions to the spin content of the octet baryons are 
reasonable. Based on them, we will do further analysis about 
the SU(3) symmetry breaking effect in HSD.

\section{Quark Distributions of Octet Baryons}

We have bulk of data to constrain the shape of  the quark distributions
of the nucleon and check various theoretical models for
the quark distributions. Starting from the  SU(6) quark
model wave function of the nucleon, the quark diquark spectator
model~\cite{Ma96,MSSY} can give the shape of valence quark distributions.
The  non-perturbative
effects such as gluon exchanges  can be effectively described
by introducing mass differences in  constituent quarks and
diquark spectators. On the other hand, in consideration of
minimally connected tree graphs of hard gluon exchanges, it was found
that  the behavior  of quark distributions at large Bjorken $x$ obeys some
counting rules~\cite{countingr} and has "helicity retention"
property~\cite{Bro95}.
Furthermore, it has been recently found that the
input-scale parton densities in the nucleon  may  be
quasi-statistical in
nature~\cite{Flambaum98,Bickerstaff90,Bhalerao96}. With a
statistical model, a vast body of polarized and unpolarized
nucleon structure functions can be well
described~\cite{Bhalerao00} even including
proper $\bar{d}-\bar{u}$ for explaining the Gottfried  sum rule.
The above three models predict  different ratios  $d(x)/u(x)$
of the nucleon,

$$
   \left\{
      \begin{array}{ll}
      (\frac{d(x)}{u(x)})_{Diquark}
      & \rightarrow 0;\\
      (\frac{d(x)}{u(x)})_{pQCD}
      & \rightarrow \frac{1}{5};\\
      (\frac{d(x)}{u(x)})_{Statistical}
      & \rightarrow 0.22.
      \end{array}
\right. $$
It is interesting that the ratio $d(x)/u(x)|_{x \to 1}=0.22$
predicted by the statistical model is
very close the pQCD prediction.
The most recent analysis~\cite{Yang99,Sch00} of experimental
data for several processes seems to support
the pQCD  and statistical predictions
of unpolarized quark flavor structure of the nucleon at $x\to 1$.
In addition, there are less free parameters in the
statistical model  than the pQCD based analysis.
All of these motivate us to extend the statistical model
from the nucleon to octet hyperons.

 Following Ref.~\cite{Bhalerao00}, the parton number density
 $d n^{IMF} /d x$ in
 the infinite-momentum frame (IMF) can be related to the
 density $d n /d E$ in the octet baryon $B$  rest frame by

 \begin{equation}
 \frac{dn^{IMF}}{d x} = \frac {M_B^2 x} {2}
 \int\limits_{x M_B/2}^
 {M_B/2} \frac{d E}{E^2} \frac{d n }{dE},\label{dnx}
 \end{equation}
where $M_B$ is the mass of the baryon $B$ and $E$ is
the parton energy in the baryon rest frame.
It should be pointed out  that Eq.~(\ref{dnx}) is
an assumption even for massless quarks since it assumes
that quarks can be boosted using a purely kinematic
transformation, which is in general not true in
an interacting theory, especially not in a strongly interacting
theory such as QCD. However, the reasonableness  of the model has
been tested by its  successful application  to the prediction of
quark distributions of the nucleon. Extending the model from the
nucleon to other members of the octet can provide an
independent check  of the same mechanism that produces the
flavor and spin structure of the nucleon and
enrich our knowledge of the nucleon.

In consideration of the finite size  effect of the baryon, 
$dn/d E$ can be expressed as
the sum of the volume, surface and curvature terms,
\begin{equation}
dn/dE= g f (E) ( V E ^2 /2 \pi^2+ a R^2 E + b R ),\label{dne}
\end{equation}
with the usual Fermi or Bose distribution function
$f(E)=1/[{\rm{e}}^{(E-\mu)/T} \pm 1 ]$.
In (\ref{dne}), $g$ is the spin-color
degeneracy factor, $V$ is the baryon  volume and $R$ is the
radius of a sphere with volume $V$. The parameters $a$ and $b$ in
(\ref{dne}) have been determined by fitting the structure function data
for the proton. We choose the same values of them for other octet  baryons,
{\it{i.e.}} $a=-0.376$ and $b=0.504$.
Then, $n_{q(\bar{q})}^{\uparrow(\downarrow)}$ which denotes the
number of quarks(antiquarks) and spin parallel
(anti-parallel) to the baryon spin can be written as

 \begin{equation}
 n_{q(\bar{q})}^{\uparrow(\downarrow)}=g \int\limits_0^{M_B/2}
 \frac{V E^2/ 2\pi ^2 + a R ^2 E +b R}
 { {\rm{e}}^{(E-\mu_{q(\bar{q})}^{\uparrow(\downarrow)})/T} + 1} dE.
 \end{equation}
Similarly, the momentum fraction carried by the quark $q$
(antiquark $\bar{q}$) and gluon $G$ can
be expressed as

 \begin{equation}
 M_{q(\bar{q})}^{\uparrow(\downarrow)}=\frac{4g}{3M_B}
 \int\limits_0^{M_B/2}
 \frac{E(V E^2/ 2\pi ^2 + a R ^2 E +b R)}
 { {\rm{e}}^{(E-\mu_{q(\bar{q})}^{\uparrow(\downarrow)})/T} + 1} dE,
 \end{equation}

 \begin{equation}
 M_{G}^{\uparrow(\downarrow)}=\frac{4g}{3 M_B}
 \int\limits_0^{M_B/2}
 \frac{E(V E^2/ 2\pi ^2 + a R ^2 E +b R)}
 { {\rm{e}}^{(E-\mu_{G}^{\uparrow(\downarrow)})/T} - 1} dE.
 \end{equation}
Hence, the quark numbers and the parton momentum fractions
have to satisfy the following 7 constraints:

 \begin{equation}
 n_q^\uparrow + n_q^\downarrow-n_{\bar{q}}^\uparrow - n_{\bar{q}}^\downarrow
=N_Q, \label{con1}
 \end{equation}

\begin{equation}
 n_q^\uparrow - n_q^\downarrow + n_{\bar{q}}^\uparrow - n_{\bar{q}}^\downarrow
 =\Delta Q, \label{con2}
 \end{equation}

\begin{equation}
\sum \limits_{q} (M_q^\uparrow + M_q^\downarrow + M_{\bar{q}}^\uparrow
+ M_{\bar{q}}^\downarrow) + (M_G^\uparrow + M_G^\downarrow)=1, \label{con5}
\end{equation}
where $q=u$, $d$ and $s$. $N_Q$ is the quark number of the baryon $B$.
In order to describe the spin structure  of the baryon,
it is necessary to distinguish between
$\mu_{q(\bar{q})}^{\uparrow}$ and $\mu_{q(\bar{q})}^{\downarrow}$.
We assume that the gluon is not polarized at the initial scale and
hence $\mu_{G}^{\uparrow}=\mu_{G}^{\downarrow}=0$. Thus at input
scale, $\Delta G(x) =0$ and the  gluon polarization comes from the
QCD evolution. In addition, it has been noticed that
$\mu_{\bar{q}}^{\uparrow} =-\mu_q^{\downarrow}$ and
$\mu_{\bar{q}}^{\downarrow} =-\mu_q^{\uparrow}$~\cite{Bhalerao00}.
Therefore, by solving 7 coupled nonlinear equations (\ref{con1})-(\ref{con5}),
we can determine 7 unknowns, namely $\mu_u^{\uparrow}$,
$\mu_u^{\downarrow}$, $\mu_d^{\uparrow}$,
$\mu_d^{\downarrow}$, $\mu_s^{\uparrow}$, $\mu_s^{\downarrow}$,
and $T$. For the $\Lambda$ and $\Sigma^0$, the 7 equations reduce into
5 equations since the $u$ and $d$ quarks in these two hyperons  are
expected to be equal due to isospin symmetry.

\begin{table*}
\caption{ Chemical potentials ($\mu$) and temperature
($T$) (in MeV) for Set-1 $\Delta Q$.}
\begin{tabular}{|c||c|c|c|c||c|c|c|c|c|c|c|}\hline
 Baryon & $\Delta U$ & $\Delta D$ & $\Delta S$ & $\Delta \Sigma$
 &$\mu_u^{\uparrow}$ &$\mu_u^{\downarrow}$
 &$\mu_d^{\uparrow}$ &$\mu_d^{\downarrow}$
 &$\mu_s^{\uparrow}$ &$\mu_s^{\downarrow}$ & $T$\\ \hline
 p & 0.82 & -0.44 & -0.10 & 0.28 & 209.6 & 87.0 & 41.0& 106.6 & -7.3 & 7.3 & 62.2\\
\hline n & -0.44 & 0.82 & -0.10 & 0.28 & 41.0 & 106.6 &209.5 &
87.0 & -7.3 & 7.3 & 62.3 \\ \hline $\Sigma^+$ & 0.82 &-0.10 &
-0.44 & 0.28 & 203.7 & 86.3 & -7.3 & 7.3 & 40.9 & 105.4 & 71.6\\
\hline $\Sigma^0$ & 0.36 & 0.36 & -0.44 & 0.28 & 99.0 & 46.6 &99.0
& 46.6 & 40.8 & 104.9 & 76.4 \\ \hline $\Sigma^-$ & -0.10 & 0.82 &
-0.44& 0.28  & -7.3 & 7.3 &203.5 & 86.3 & 40.9 & 105.4 & 71.9\\
\hline $\Lambda^0$ & -0.17 & -0.17 & 0.62 & 0.28 & 60.7 & 85.6
&60.7 & 85.6 & 118.5 & 27.7 & 74.0\\ \hline $\Xi^-$ & -0.10 &
-0.44 & 0.82& 0.28 & -7.3 & 7.3 &40.7 & 104.6 & 201.1 & 85.8 &
75.9\\ \hline $\Xi^0$ & -0.44 & -0.10 & 0.82& 0.28  & 40.7 & 104.7
&-7.3 & 7.3 & 201.2 & 85.8 & 75.7\\ \hline
\end{tabular}\label{table2}
\end{table*}

\begin{table*}
\caption{ Chemical potentials ($\mu$) and temperature
($T$) (in MeV) for Set-2 $\Delta Q$. }
\begin{tabular}{|c||c|c|c|c||c|c|c|c|c|c|c|}\hline
 Baryon & $\Delta U$ & $\Delta D$ & $\Delta S$ & $\Delta \Sigma$
 &$\mu_u^{\uparrow}$ &$\mu_u^{\downarrow}$
 &$\mu_d^{\uparrow}$ &$\mu_d^{\downarrow}$
 &$\mu_s^{\uparrow}$ &$\mu_s^{\downarrow}$ & $T$\\ \hline
 p & 0.78 & -0.48 & -0.22 & 0.08 & 206.7 & 90.0 & 38.1 & 109.6 & -16.1 & 16.1 & 62.2\\
\hline n & -0.48 & 0.78 & -0.22 & 0.08 & 38.1 & 109.6 & 206.7 &
90.0 & -16.1 & 16.1 & 62.3 \\ \hline $\Sigma^+$ & 0.56 &-0.18 &
-0.60& -0.22 & 185.8 & 105.4 &-13.1 & 13.1 & 29.2 & 117.1 & 72.0\\
\hline $\Sigma^0$ & 0.19 & 0.19 & -0.60 & -0.22 & 86.7 & 59.0
& 86.7 & 59.0 & 29.1 & 116.4 & 76.4 \\ \hline $\Sigma^-$ & -0.18 &
0.56 & -0.60 & -0.22 & -13.1 & 13.1 & 185.7 & 105.3 & 29.2 & 117.0
& 72.3 \\ \hline $\Lambda^0$ & 0.03 & 0.03 & 0.65 & 0.71 & 75.3 &
70.9 & 75.3 & 70.9 & 120.7 & 25.6 & 74.1\\ \hline $\Xi^-$ & 0.08 &
-0.13 & 0.91 & 0.86 & 5.8 & -5.8 & 63.2 & 82.2 & 207.0 & 79.2 &
75.9 \\ \hline $\Xi^0$ & -0.13 & 0.08 & 0.91 & 0.86 & 63.3 & 82.2
& 5.8 & -5.8 & 207.1 & 79.3 & 75.7 \\ \hline
\end{tabular}\label{table3}
\end{table*}

The important inputs for the statistical model  are  quark contributions
to the spin content of a baryon.
We have noticed that the  SU(3) flavor  symmetry breaking in
HSD has a significant effect on the  extraction of the quark
contribution $\Delta Q$  to the spin of
the octet baryons. In order to check the effect  of SU(3)
symmetry breaking, we adopt two sets of typical $\Delta Q$:
Set-1 corresponds to the SU(3) symmetry
case with the same $\Delta \Sigma=0.28$ for all octet
baryons~\cite{Bor98};
Set-2 corresponds to the SU(3) broken case with
$\Delta Q$ obtained in the last section.
 The corresponding solutions of
$\mu_q^{\uparrow}$, $\mu_q^{\downarrow}$ ($q=u$, $d$, $s$) and $T$
for Set-1 and Set-2 $\Delta Q$'s are listed in Table~\ref{table2} and 
Table~\ref{table3}, respectively. With these values of the chemical potentials
and temperature,  unpolarized and
polarized parton distributions  in the octet baryons
can be obtained directly from (\ref{dnx}).

We find that the SU(3)
symmetry breaking effect on quark spin
structures of the octet hyperons is significant.
Therefore, the octet hyperons are
suitable  laboratories to examine SU(3) symmetry breaking in HSD.

\section{Octet Baryon Polarizations from Fragmentation}

\begin{figure*}
\includegraphics[width=10cm,height=16cm]{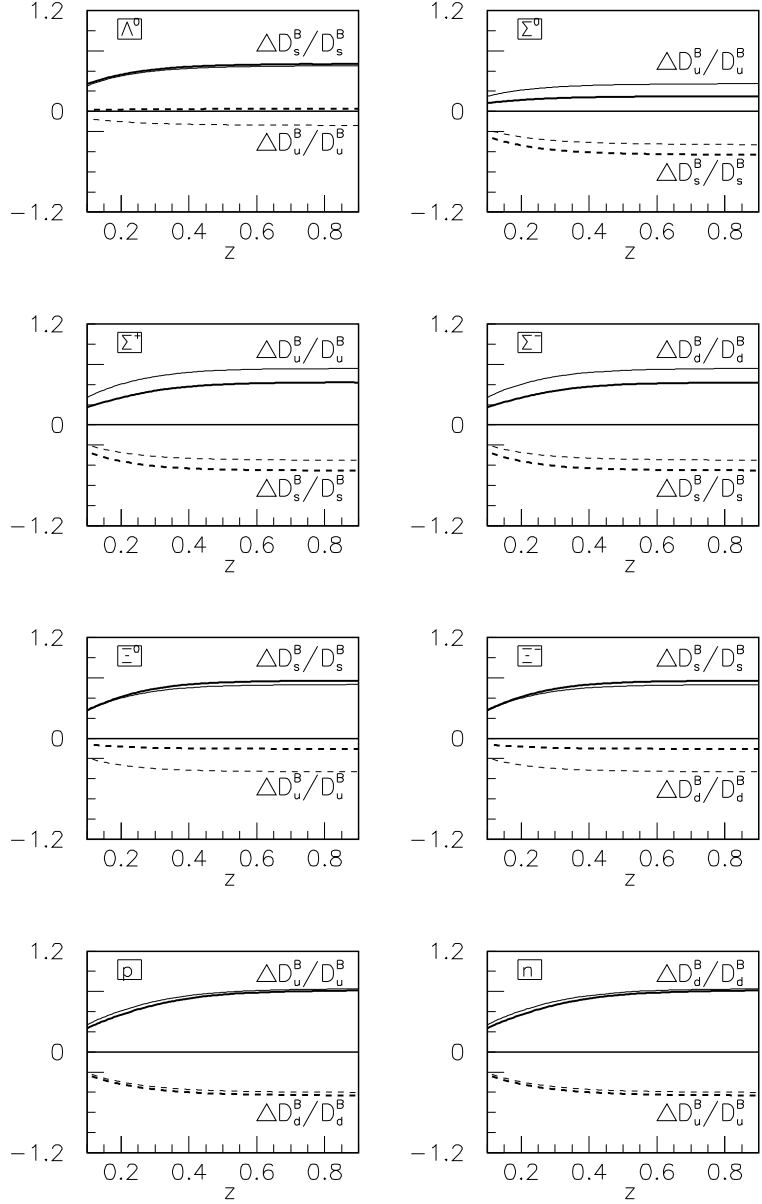}
\caption[*]{\baselineskip 13pt The spin structure of the
fragmentation functions for the octet baryons. The solid and dashed
 curves are for the dominant and non-dominant quarks, respectively.
 The thin and thick curves correspond to the Set-1 and Set-2
 fragmentation functions. }\label{a04f1}
\end{figure*}

Unfortunately, we cannot check the SU(3) symmetry  breaking
effect on the obtained quark  distributions of the octet hyperons
by means of structure functions in DIS scattering since they
cannot be used as a target due to their  short life time. Also one
obviously cannot produce a beam of charge-neutral hyperons such as
$\Lambda$. For this reason, some efforts have been made to model
fragmentation functions for the $\Lambda$~\cite{Nza95, Bor99,
Yang02}. On the other hand, there have been  attempts to connect
the quark distributions with the quark fragmentation functions, so
that one can explore the quark structure of hyperons 
by means of hyperon production from quark fragmentation.
The connection is the so
called Gribov-Lipatov (GL) relation~\cite{GLR}
\begin{equation}
D_q^h(z) \sim z q_h(z),
\label{GLR}
\end{equation}
where $D_q^h(z)$ is the fragmentation function for a
quark $q$ splitting into a hadron $h$ with longitudinal
momentum fraction $z$, and $q_h(z)$ is the quark
distribution of finding the quark $q$ carrying a
momentum fraction $x=z$ inside the hadron $h$. The GL relation should
be considered as an approximate relation near $z \to 1$
on an energy scale $Q^2_0$~\cite{BRV00,Bar00}.
It is interesting to note that such a relation provided
successful descriptions of the available $\Lambda$ polarization
data in several processes~\cite{MSSY,Yang01},
based on quark distributions of the $\Lambda$ in
 the quark diquark model, pQCD based counting rules
analysis and statistical model.
Thus we still use (\ref{GLR}) as an Ansatz to relate
the quark fragmentation functions for the
octet baryons  to the corresponding quark distributions
at an initial scale. Then, the quark fragmentation functions
are evolved from the initial scale to the experimental
energy scale. We use the evolution package of
Ref.~\cite{Miyama94} suitable modified for the evolution of
fragmentation functions in leading order, taking
the initial scale $Q_0^2=M_B^2$ and $\Lambda_{QCD}=0.3~\rm{GeV}$.
In Fig.~\ref{a04f1}, the spin  properties  of
the Set-1 (thin curves) and Set-2 (thick curves)
fragmentation functions are presented at $Q^2=4~\rm{GeV}^2$.
From Fig.~\ref{a04f1}, we find that the difference between the two sets
of fragmentation functions for the nucleon is very small.
However, the SU(3) symmetry breaking effect on the
fragmentation functions for the octet hyperons
is significant and it might be  detected via
hyperon production. For this reason, we use the obtained
fragmentation functions to predict spin observables
for the various hyperon production processes.

We need some experimental data to examine the SU(3) 
symmetry breaking effect on the quark spin structure 
of the octet baryons. There has been some
recent progress in the measurements of polarized $\Lambda$
production. The longitudinal $\Lambda$ polarization in $e^+e^-$
annihilation at the Z-pole has been measured  by several
collaborations~\cite{ALEPH96,DELPHI95,OPAL97}. The HERMES
Collaboration at DESY~\cite{HERMES} and the E665 Collaboration at
FNAL~\cite{E665}  reported their results for the longitudinal spin
transfer to the $\Lambda$ in polarized positron DIS.
Very recently, the measurement results  of $\Lambda$ polarization
in charged current interactions were obtained by
the NOMAD Collaboration~\cite{NOMAD}. We can check whether 
the SU(3) symmetry breaking effect in $\Lambda$ 
production is supported by the available 
experimental data. In order to obtain some more information on 
SU(3) symmetry breaking, we also predict polarizations for other octet
baryons produced in $e^+e^-$ annihilation, charged lepton
DIS and neutrino DIS.

\subsection{Octet baryon polarization in $e^+e^-$ annihilation}

\begin{figure*}
\includegraphics[width=10cm,height=16cm]{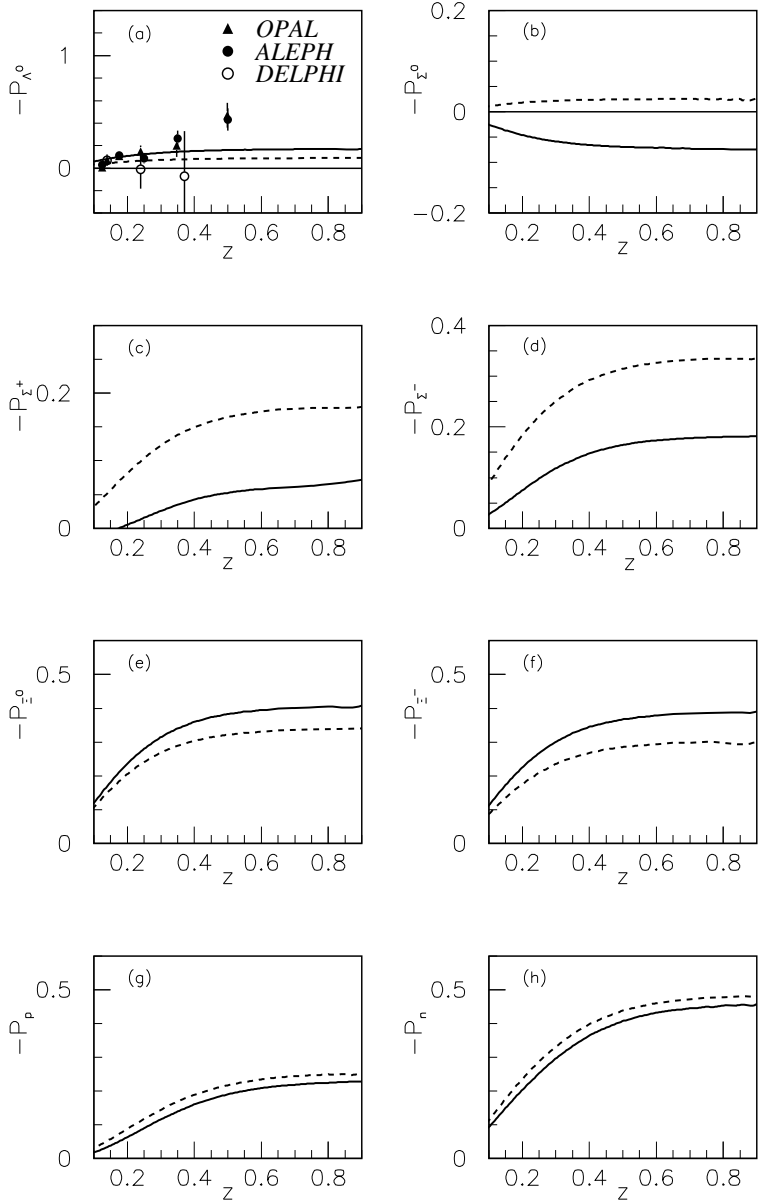}
\caption[*]{\baselineskip 13pt The longitudinal octet baryon
polarizations $P_{B}$ in $e^+e^-$ annihilation at the Z-pole.
The dashed and solid  curves are the predictions using the
Set-1 and Set-2 fragmentation functions for the octet
baryons, respectively.
The experimental data are taken from
Refs.~\cite{ALEPH96,DELPHI95,OPAL97}.}\label{a04f2}
\end{figure*}

Within the framework of the standard model of electroweak interactions,
the qaurks and antiquarks produced in $e^+e^-$-annihilation
near the $Z$-pole are polarized due to the interference between
the vector and axial vector couplings, even though
the initial $e^+$ and $e^-$ beams are unpolarized. Then the
polarized quarks and antiquarks lead to the polarization
of a baryon produced from fragmentation. Theoretically,
the baryon (B) polarization can be expressed as

\begin{equation}
P_{B}=-\frac{\sum\limits_{q} \hat{A}_q [\Delta D_q^B
(z)-\Delta D_{\bar q}^B (z)]}{\sum\limits_{q} \hat{C}_q
[D_q^B (z)+ D_{\bar q}^B (z)]}, \label{PL2}
\end{equation}
where $\hat{A}_q$ and $\hat{C}_q$ ($q=u, d$ and $s$) can be found
in Refs.~\cite{MSSY,Yang02}, and $D_q^B(z)$ ($D_{\bar{q}}^B(z)$) and
$\Delta D_q^B(z)$ ($\Delta D_{\bar{q}}^B(z)$)  are the unpolarized
and polarized quark $q$ (antiquark ${\bar{q}}$) to baryon $B$
fragmentation functions. With the obtained two sets of
fragmentation functions, we can check whether there exist some
SU(3) symmetry breaking effects on the baryon polarizations. Our
theoretical predictions for the octet baryon polarizations at the
$Z$-pole are shown in Fig.~\ref{a04f2}.  We find from
Fig.~\ref{a04f2}(a) that the prediction with SU(3) symmetry
breaking is closer to the experimental data than the prediction with SU(3)
symmetry although both predictions are only qualitatively 
compatible with the experimental data. From Fig.~\ref{a04f2}, 
one can see that the SU(3) symmetry breaking effect on the 
octet hyperon polarizations  is much more significant than on 
the nucleon polarizations. For  $\Sigma^0$
production, the SU(3) symmetry breaking effect leads to  changes
in  the sign of the polarization. The effect is also obvious for
$\Sigma^{\pm}$ production. Therefore,  the SU(3) symmetry breaking
effect might be observed via hyperon production 
in $e^+e^-$ annihilation, especially for $\Sigma$ production.

\subsection{Spin transfer to octet baryon in charged lepton DIS}

\begin{figure*}
\includegraphics[width=10cm,height=16cm]{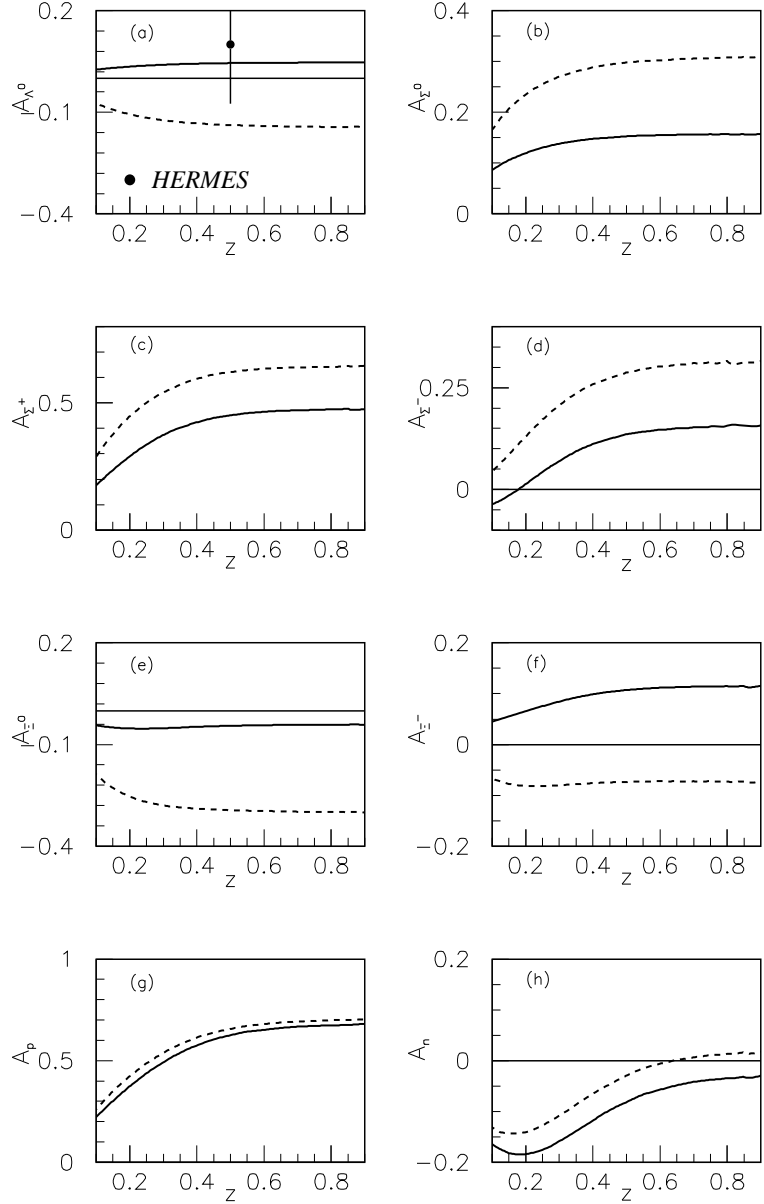}
\caption[*]{\baselineskip 13pt The predictions of the
$z$-dependence for the spin transfers to the octet baryons
in polarized charged lepton DIS on the proton target.
The dashed and solid  curves are
the predictions using the
Set-1 and Set-2 fragmentation functions for the octet
baryons, respectively. The HERMES data is taken from Ref.~\cite{HERMES}.
}\label{a04f3}
\end{figure*}

In the deep inelastic scattering of a longitudinally polarized
charged lepton on an unpolarized nucleon target,
the polarization of the beam leads  struck quark to
be polarized and its spin will be transferred to a baryon
produced via the quark fragmentation.  The
longitudinal spin transfer to the produced baryon  $B$ is
given in the quark-parton model by
\begin{equation}
A_{B}(x,z)= \frac{\sum\limits_{q} e_q^2 [q^N(x,Q^2) \Delta
D_q^B(z,Q^2) + ( q \rightarrow \bar q)]}{\sum\limits_{q} e_q^2
[q^N (x,Q^2) D^B_q(z,Q^2) + ( q \rightarrow \bar q)]}~,
\label{DL}
\end{equation}
where $q^N(x,Q^2)$ is the quark distribution of the target nucleon
and will be adopted as  the CTEQ5 set 1 parametrization
form~\cite{CTEQ5} in our numerical calculation.
The spin transfers to the octet baryons can be calculated
with the two sets of fragmentation functions and the results
are shown in Fig.~\ref{a04f3}.

From Fig.~\ref{a04f3}. We find that the
SU(3) symmetry breaking effect on the spin transfers to the octet
hyperons is much stronger than on the spin transfer to the
nucleon. For the $\Lambda$ and  $\Xi^-$ hyperons, the sign of the
spin transfer is changed  due to the SU(3) symmetry breaking effect. The
reason for this big change lies in  that the SU(3) symmetry breaking
effect leads to the positively polarized $u$ quark fragmentation
function whereas the SU(3) symmetry gives the negatively polarized $u$
quark fragmentation function (cf. Tables~\ref{table2} and \ref{table3}).
 There has been preliminary data
by HERMES Collaboration~\cite{HERMES} on $\Lambda$ production.
It is interesting that the SU(3) symmetry breaking effect
on the spin transfer to $\Lambda$ makes the prediction to be
consistent with the HERMES data point, which indicates
that inclusion of the SU(3) symmetry breaking effect improves  
the spin structure of the $\Lambda$. The modifications for other hyperons
due to SU(3) symmetry breaking
are  also significant.  Therefore, the spin transfers to the hyperons in
charged lepton DIS  are  another set of suitable observables
to check the existence of SU(3) symmetry breaking in HSD.
 Although the present data are not
sufficient to draw  a quantitative conclusion, it seems that the data favor
somewhat the  case in which the SU(3) symmetry breaking effect works.

\subsection{Baryon polarizations in neutrino DIS}

The neutrino DIS scattering on the nucleon target can
also provide a source of polarized quarks and
should be another important process to study the
polarizations of the octet baryons.
For baryon $B$ production in the current
fragmentation region from neutrino
and antineutrino DIS, the longitudinal polarization of $B$
in its momentum direction can be expressed as~\cite{Ma99},

\begin{widetext}
\begin{equation}
P_\nu^B(x,y,z)=-\frac{[d^N(x)+\varpi s^N(x)] \Delta D _u^B (z) -( 1-y)
^2 \bar{u}^N (x) [\Delta D _{\bar{d}}^B (z)+\varpi \Delta
D_{\bar{s}}^B(z)]} {[d^N(x)+\varpi s^N(x)] D_u ^B (z) + (1-y)^2
\bar{u}^N (x) [D _{\bar{d}}^B (z)+\varpi D_{\bar{s}}^B(z)]}~,
\end{equation}

\begin{equation}
P_{\bar{\nu}}^B (x,y,z)=-\frac{( 1-y) ^2 u^N (x) [\Delta D _d^B
(z)+\varpi \Delta D _s^B (z)]-[\bar{d}^N(x)+\varpi \bar{s}^N(x)]
\Delta D _{\bar{u}}^B (z)}{(1-y)^2 u^N (x) [D _d^B (z)+\varpi D _s^B
(z)]+[\bar{d}^N(x)+\varpi \bar{s}^N(x)] D_{\bar{u}}^B (z)}~,
\end{equation}
\end{widetext}
where the terms with the factor $\varpi=\sin^2 \theta_c/\cos^2
\theta_c$ ($\theta_c$ is the Cabibbo angle) represent Cabibbo
suppressed contributions. The beam can be either neutrino or
antineutrino, and the produced hadron can be either baryon or
antibaryon. Therefore, we have four combinations of different
beams and produced baryons and can get rich information on the
flavor dependence of quark fragmentation functions.

\begin{figure*}
\includegraphics[width=10cm,height=8cm]{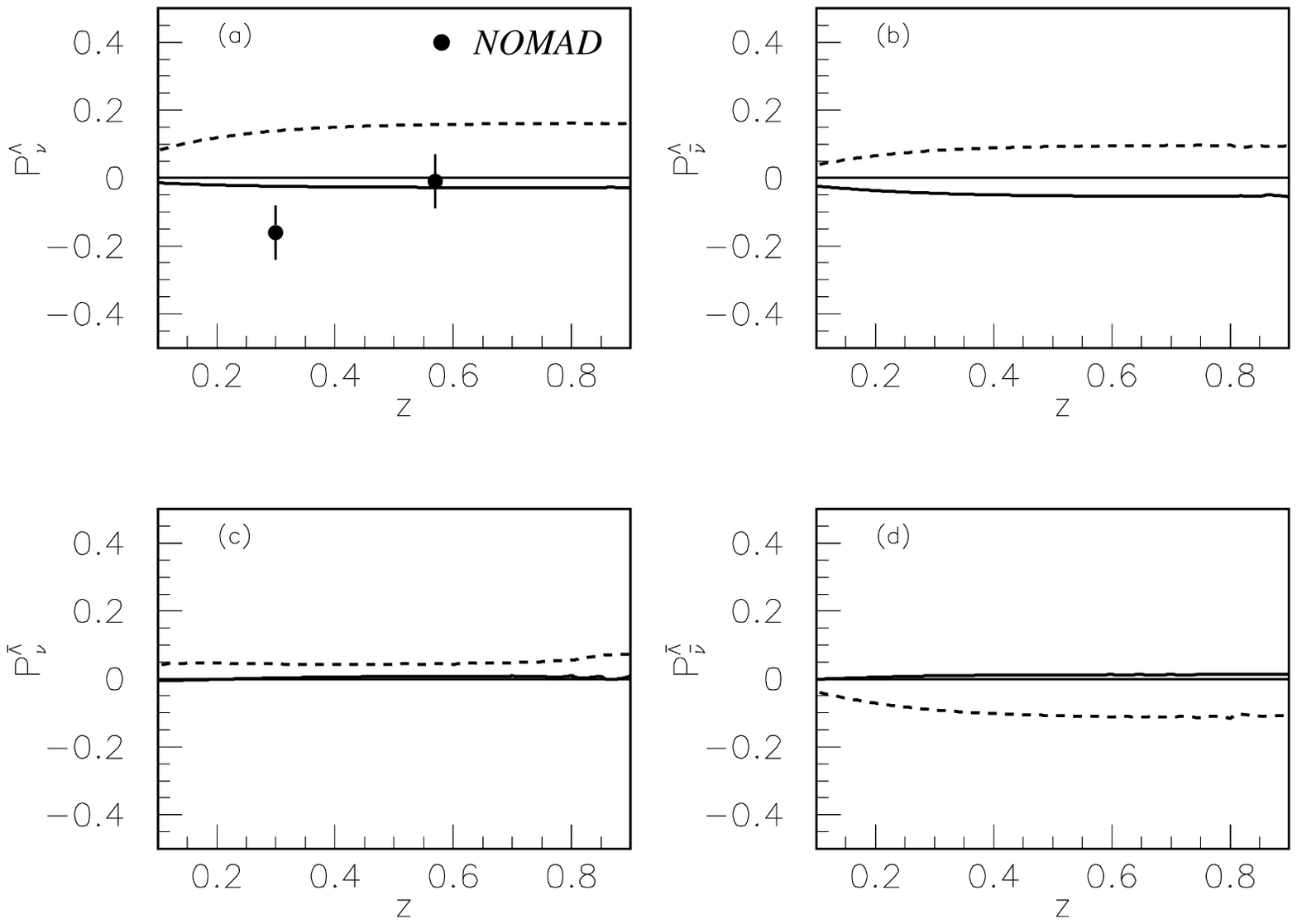}
\caption[*]{\baselineskip 13pt The predictions of $z$-dependence
for the hadron and anti-hadron polarizations of $\Lambda$ in
neutrino (antineutrino) DIS. The dashed and solid curves
correspond to the predictions by using the Set-1 and Set-2
fragmentation functions with the Bjorken variable $x$
integrated over $0.02 \to 0.4$ and $y$ integrated
over $0 \to 1$. }\label{a04f4}
\end{figure*}

\begin{figure*}
\includegraphics[width=10cm,height=8cm]{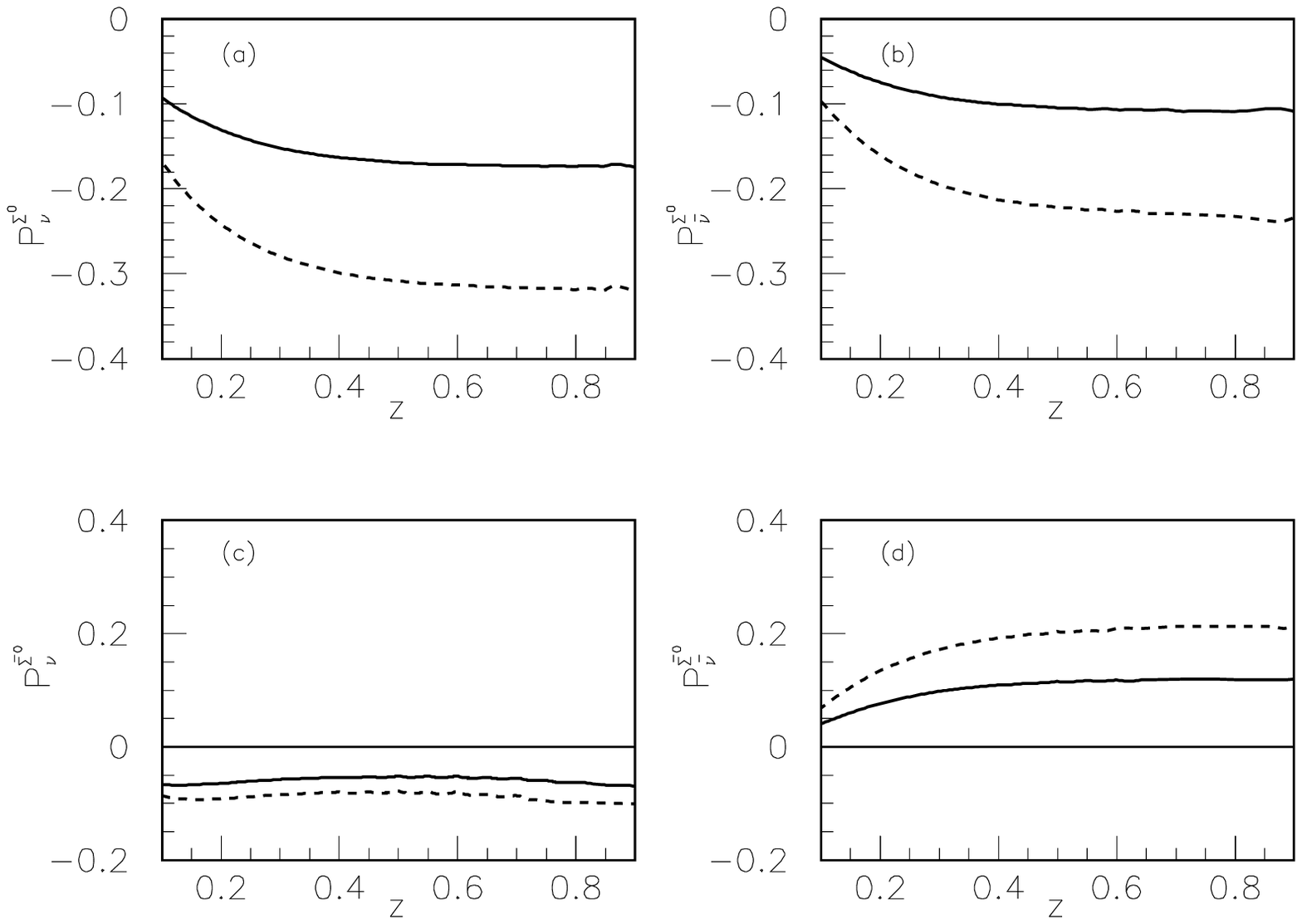}
\caption[*]{\baselineskip 13pt The same as Fig.~\ref{a04f4},
but for predictions of $z$-dependence for the hadron and
anti-hadron polarizations of $\Sigma^0$ in neutrino
(antineutrino) DIS. }\label{a04f5}
\end{figure*}

\begin{figure*}
\includegraphics[width=10cm,height=8cm]{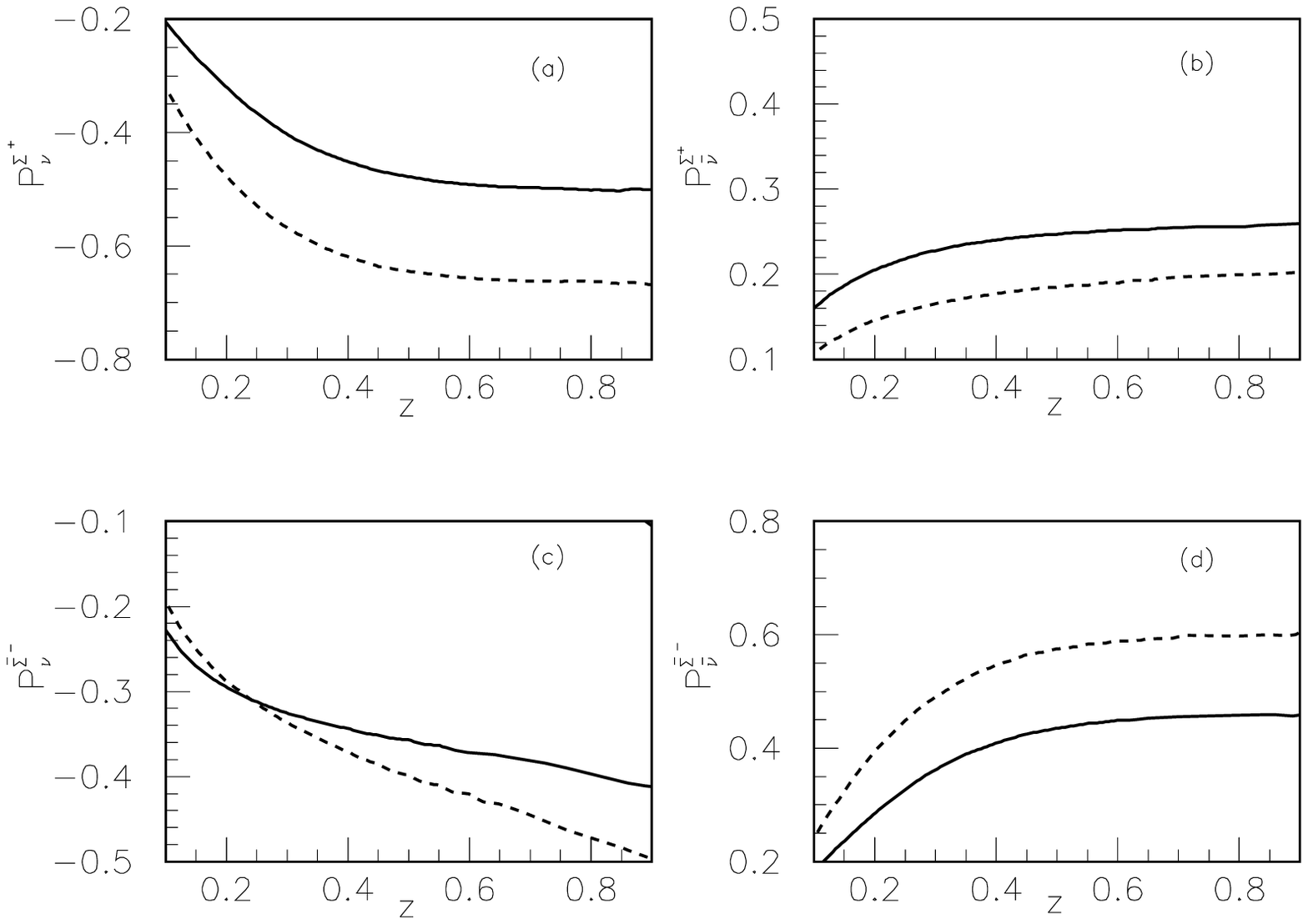}
\caption[*]{\baselineskip 13pt The same as Fig.~\ref{a04f4},
but for predictions of $z$-dependence for the hadron and
anti-hadron polarizations of $\Sigma^+$ in neutrino
(antineutrino) DIS. }\label{a04f6}
\end{figure*}

\begin{figure*}
\includegraphics[width=10cm,height=8cm]{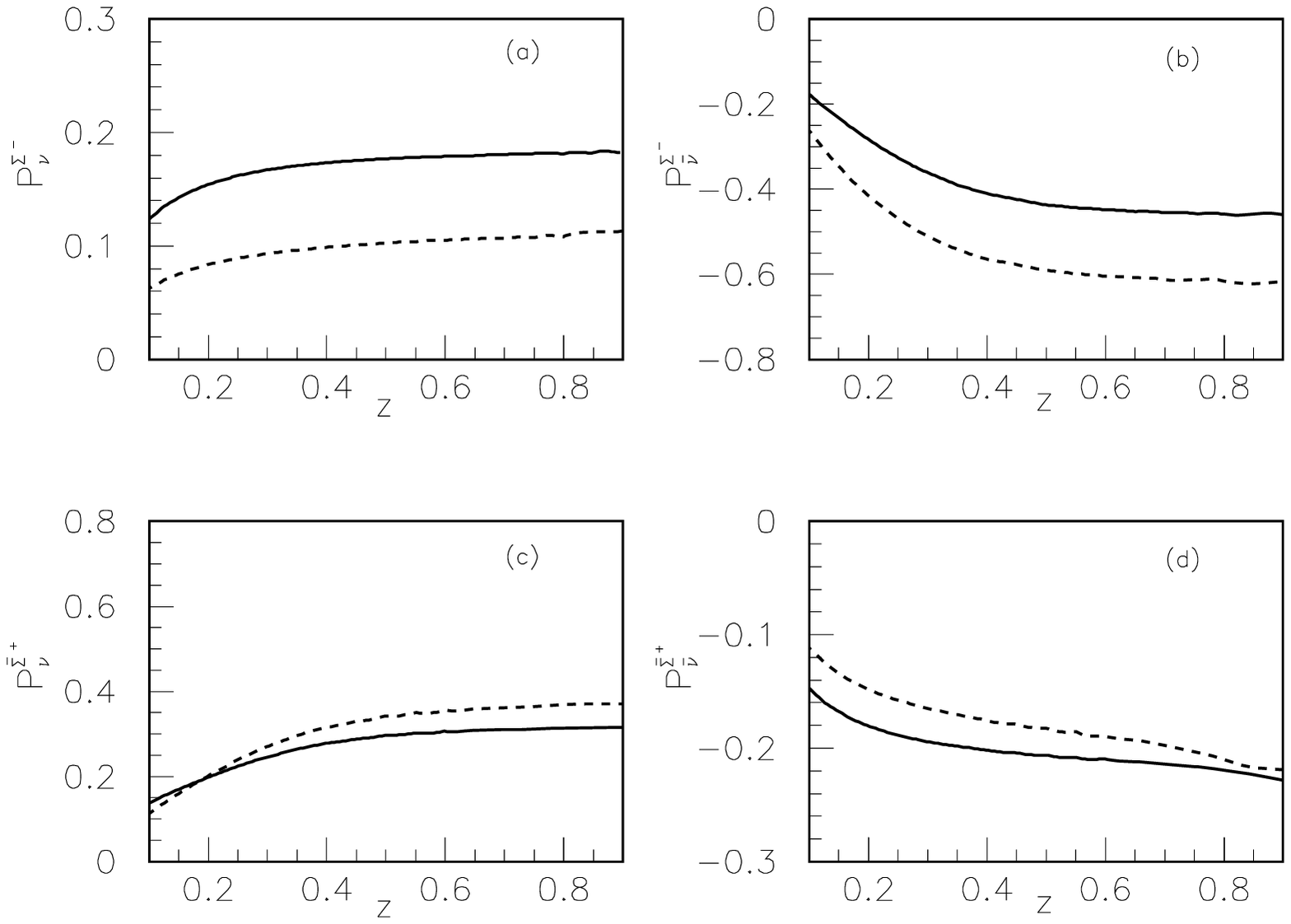}
\caption[*]{\baselineskip 13pt The same as Fig.~\ref{a04f4}, but
for predictions of $z$-dependence for the hadron and anti-hadron
polarizations of $\Sigma^-$ in  neutrino (antineutrino) DIS.
}\label{a04f7}
\end{figure*}

\begin{figure*}
\includegraphics[width=10cm,height=8cm]{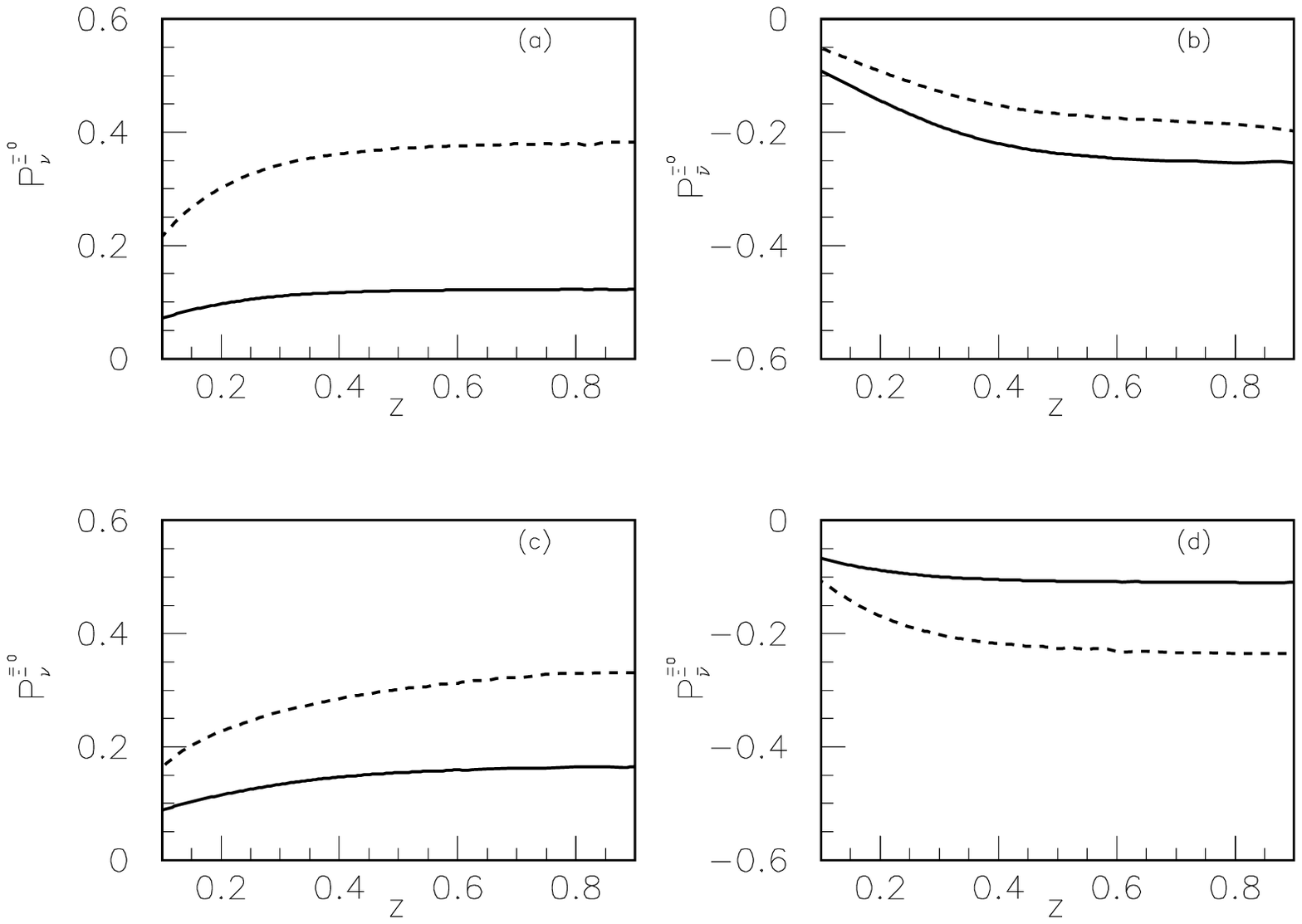}
\caption[*]{\baselineskip 13pt The same as Fig.~\ref{a04f4},
but for predictions of $z$-dependence for the hadron and
anti-hadron polarizations of $\Xi^0$ in neutrino
(antineutrino) DIS. }\label{a04f8}
\end{figure*}

\begin{figure*}
\includegraphics[width=10cm,height=8cm]{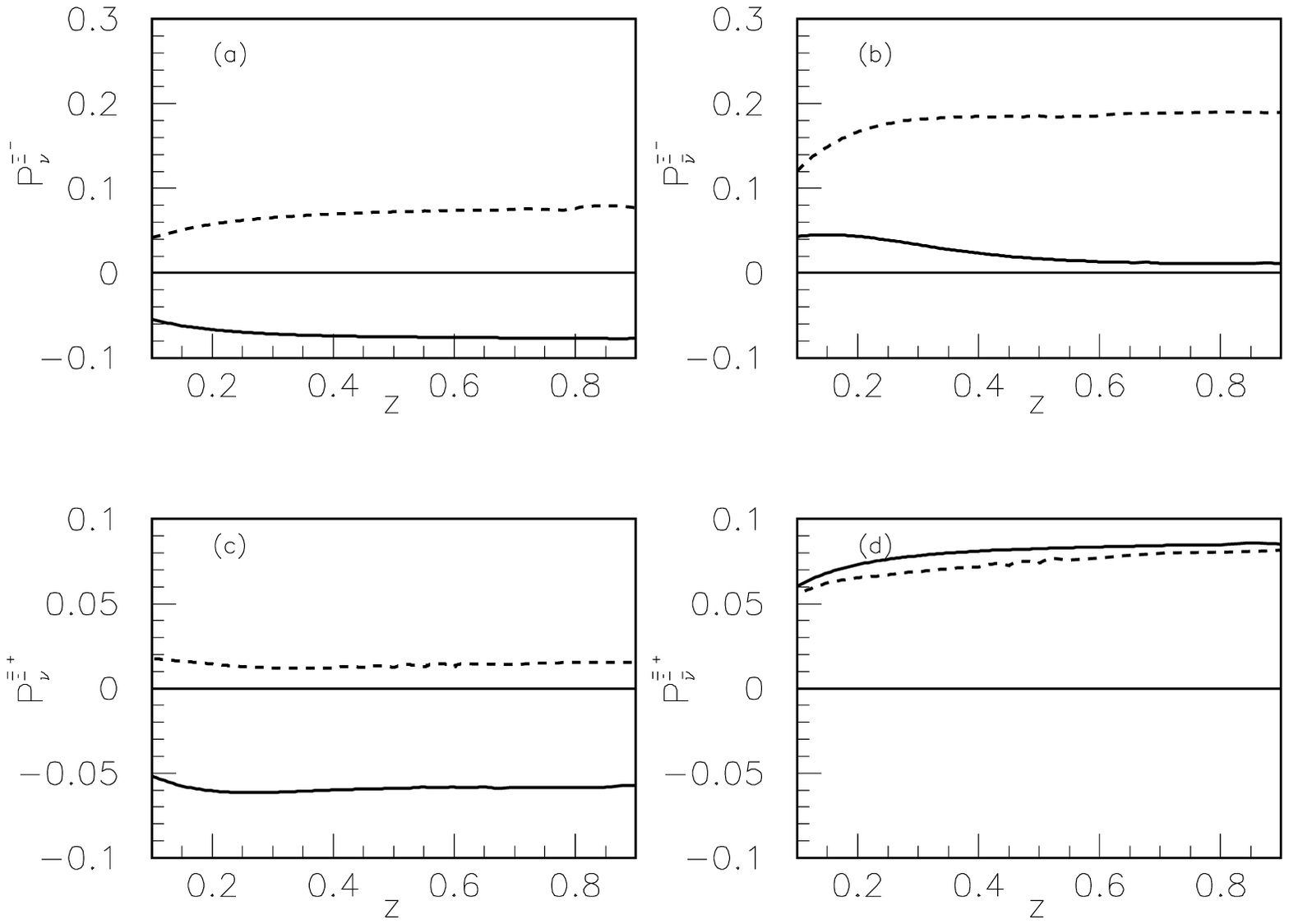}
\caption[*]{\baselineskip 13pt The same as Fig.~\ref{a04f4},
but for predictions of $z$-dependence for the hadron and
anti-hadron polarizations of $\Xi^-$ in neutrino
(antineutrino) DIS. }\label{a04f9}
\end{figure*}

\begin{figure*}
\includegraphics[width=10cm,height=8cm]{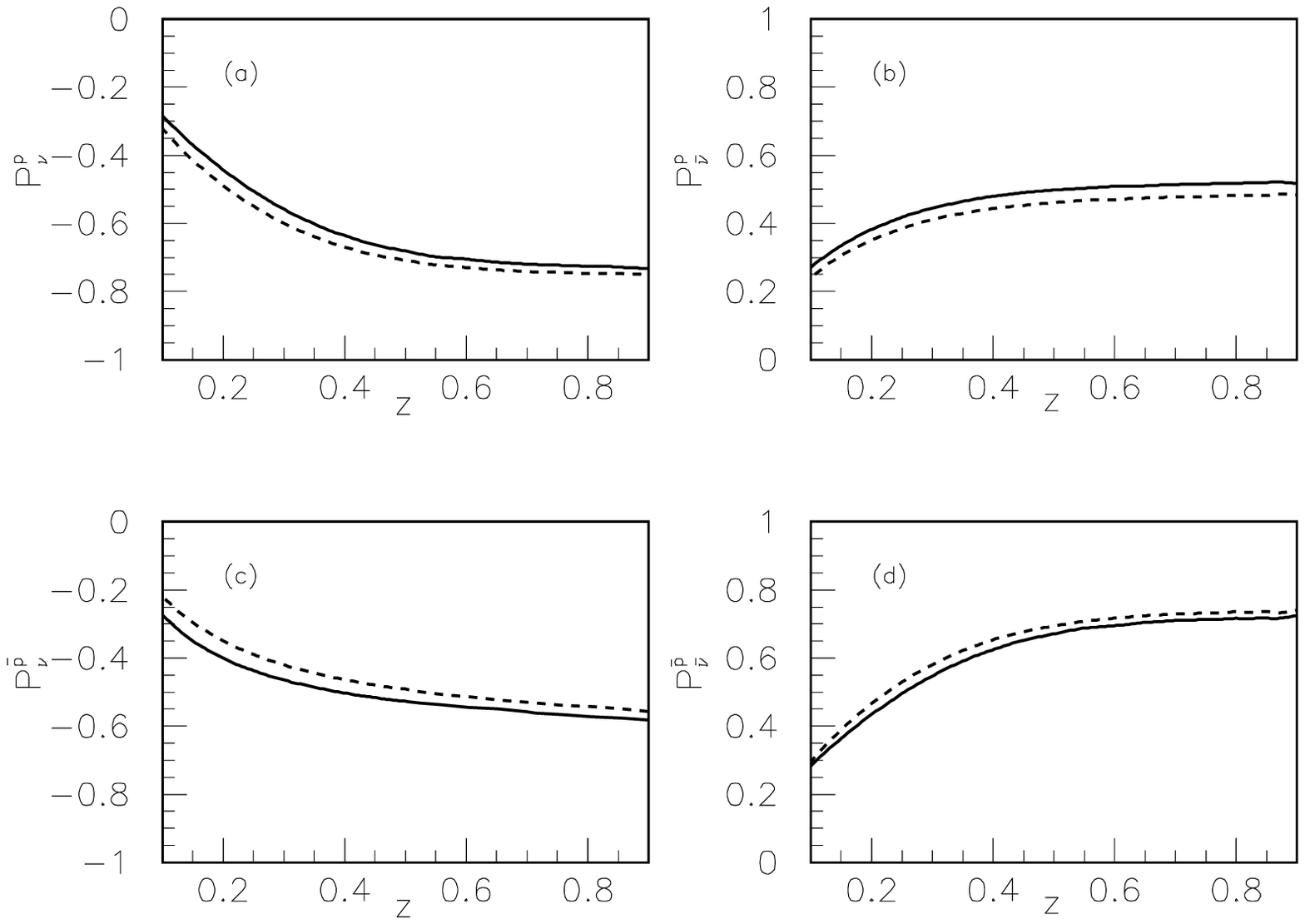}
\caption[*]{\baselineskip 13pt The same as Fig.~\ref{a04f4},
but for predictions of $z$-dependence for the hadron and
anti-hadron polarizations of $p$ in neutrino (antineutrino)
DIS. }\label{a04f10}
\end{figure*}

\begin{figure*}
\includegraphics[width=10cm,height=8cm]{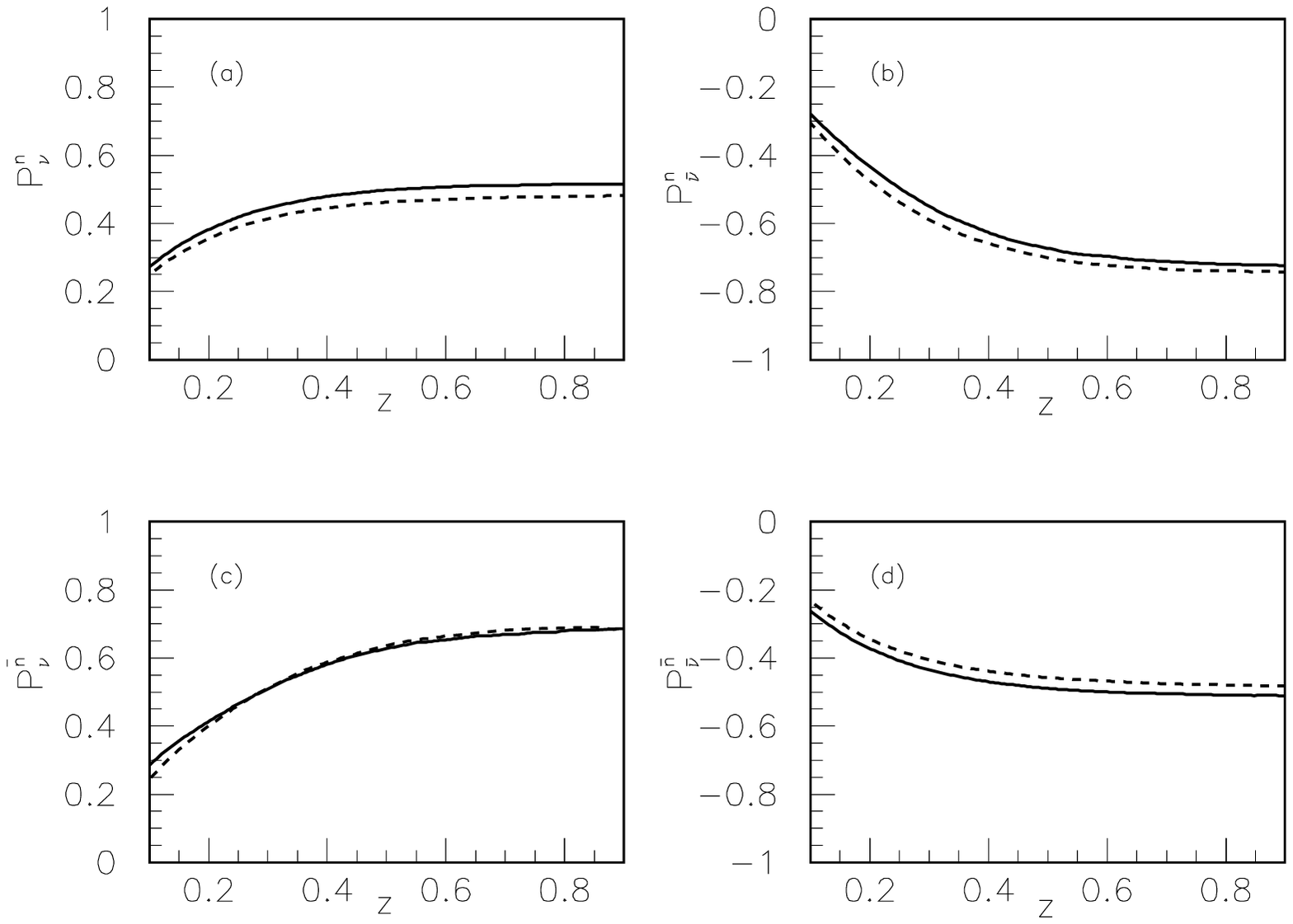}
\caption[*]{\baselineskip 13pt The same as Fig.~\ref{a04f4},
but for predictions of $z$-dependence for the hadron and
anti-hadron polarizations of $n$ in neutrino (antineutrino)
DIS. }\label{a04f11}
\end{figure*}

We present in Figs.~\ref{a04f4}-\ref{a04f11} the longitudinal
polarizations for the octet baryons produced
in neutrino (antineutrino) DIS. For $\Lambda$ production, 
there have been preliminary experimental data by NOMAD
Collaboration~\cite{NOMAD}. The NOMAD data have
relative high precision and can be used to distinguish different
predictions. The data seem to support again the Set-2
prediction with  SU(3) symmetry breaking. The polarizations 
$P_{\nu}^{\Lambda}$, $P_{\bar{\nu}}^{\Lambda}$, $P_{\nu}^{\Xi^-}$ and 
$P_{\nu}^{\bar{\Xi}^+}$ are very valuable for us to reveal  
the SU(3) symmetry breaking effect.
 The  octet hyperon polarizations are more sensitive
 to SU(3) symmetry breaking than the nucleon polarizations.
We can also obtain some useful information on SU(3) symmetry breaking
by means of the polarizations of the octet hyperons
produced in neutrino (antineutrino) DIS.

\section{Summary and Discussion}

Based on the known data of the semileptonic decays
and lepton-nucleon deep inelastic scattering, we
re-extracted quark contributions to the spin content of the octet
baryons with  SU(3) symmetry breaking in order to get reasonable 
central values for them. We constrained the quark distributions  of the
octet baryons  at an initial scale with the two sets
of typical quark contributions to the spin content of the octet baryons:
one set with  SU(3) flavor  symmetry and another
set with  SU(3)  flavor symmetry breaking. By means of the
statistical model, we calculated quark distributions
for the octet baryons in the rest frame and then used 
free boost transformations
to relate the rest frame results to the IMF and
made predictions about PDFs. We find that quark distributions
of the octet hyperons are much more sensitive to SU(3) symmetry breaking
 than those of the nucleon.
In consideration of the fact that it is difficult for one to access
the SU(3) symmetry breaking effect on the quark
distributions of the octet hyperons,
we focused  our attention on exploring  the possible SU(3) symmetry 
breaking effect on the octet baryon polarization from fragmentation.
It was found that the available experimental data on 
$\Lambda$ production seem to favor the predictions with SU(3)
symmetry breaking. The spin observables in hyperon production
from quark fragmentation, especially the polarization of the $\Sigma$ in 
$e^+e^-$ annihilation, the spin transfers in charged lepton DIS and 
polarizations in neutrino DIS for the $\Lambda$ and $\Xi$ hyperons, 
are very valuable for us to reveal the SU(3) symmetry breaking effect. 
We find that the high precision measurement on the
hyperon polarizations  in $e^+e^-$ annihilation,
charged lepton DIS and neutrino DIS  can provide  a possible 
collateral evidence for the SU(3)
symmetry breaking in HSD of the octet baryons.

\begin{figure*}
\includegraphics[width=10cm,height=10cm]{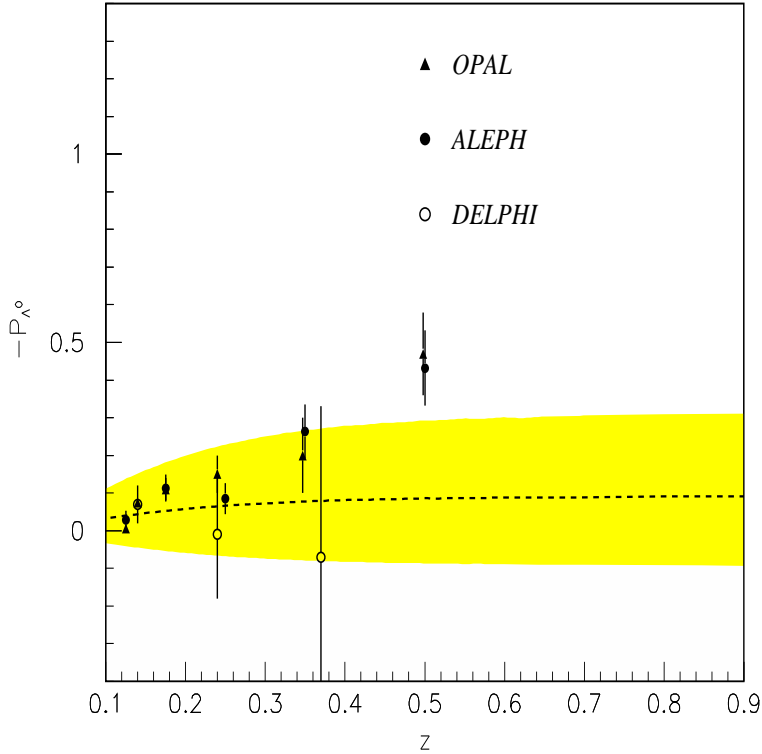}
\caption[*]{\baselineskip 13pt The same as Fig.~\ref{a04f2}(a),
but the single solid curve has been replaced by a band due to the consideration of errors 
on $\Delta Q$'s ($Q=U$, $D$ and $S$). The dashed curve is for 
the SU(3) symmetry case.}\label{a04f12}
\end{figure*}

\begin{figure*}
\includegraphics[width=10cm,height=10cm]{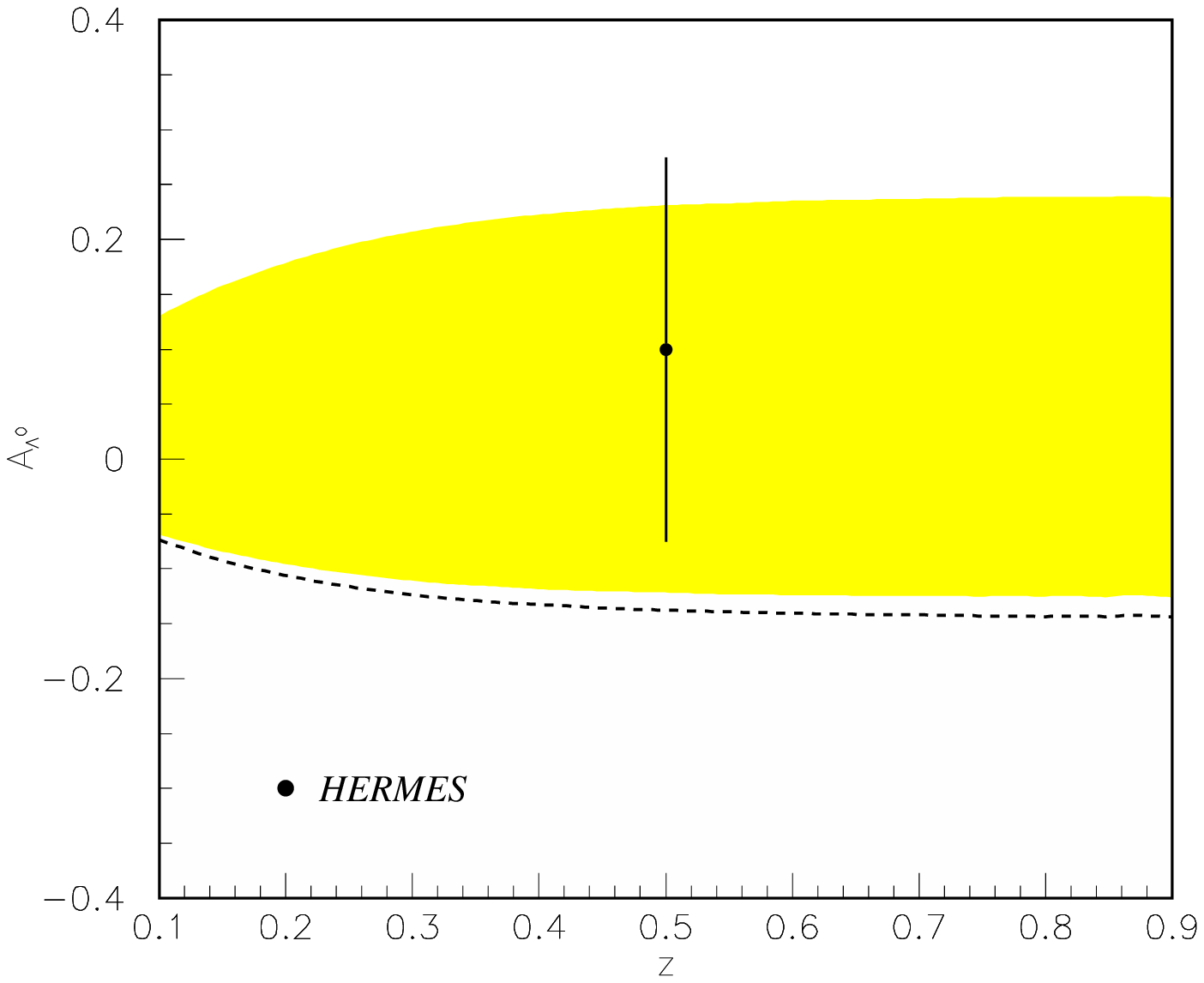}
\caption[*]{\baselineskip 13pt The same as Fig.~\ref{a04f3}(a),
but the single solid curve has been replaced by a band due to 
the consideration of errors on $\Delta Q$'s ($Q=U$, $D$ and $S$). 
The dashed curve is for the SU(3) symmetry 
case.}\label{a04f13}
\end{figure*}

\begin{figure*}
\includegraphics[width=10cm,height=10cm]{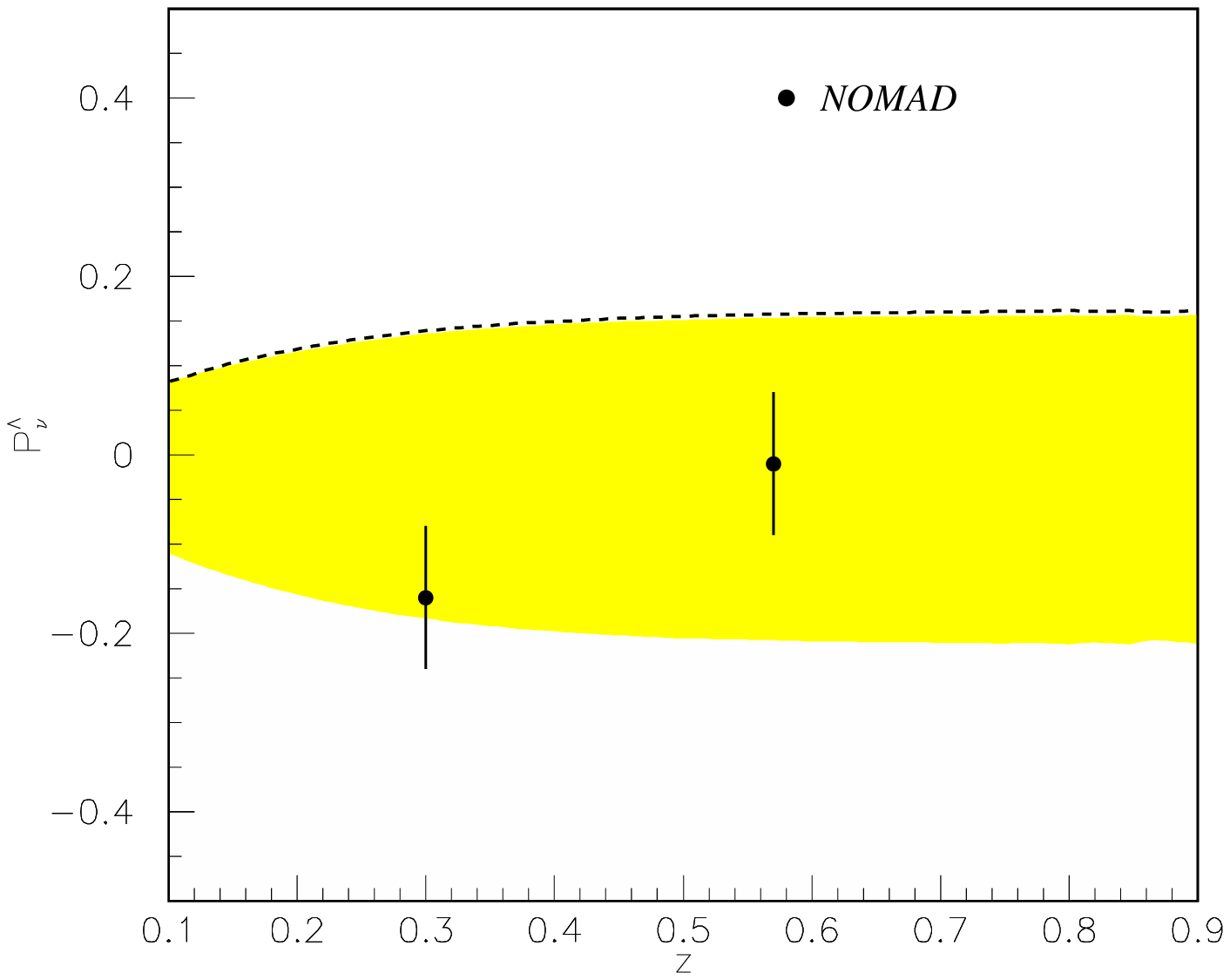}
\caption[*]{\baselineskip 13pt The same as Fig.~\ref{a04f4}(a),
but the single solid curve has been replaced by a band due to 
the consideration of errors on $\Delta Q$'s ($Q=U$, $D$ and $S$). 
The dashed curve is for the SU(3) symmetry 
case.}\label{a04f14}
\end{figure*}

It is worth to mention that there exist many uncertainties in our predictions
and there are many unknowns to be explored before we can arrive at 
a firm conclusion. We used the following three model assumptions:
(1) an approximate relation between  the fragmentation functions
and quark distributions functions, Eq.(\ref{GLR}); (2) the model for the 
symmetry breaking, Refs.~\cite{KimPRD, Manohar98, KimNPA}; (3) 
the model for the parton densities, Ref.~\cite{Bhalerao00}. 
Each of these assumptions introduces its own uncertainties into 
our predictions. It should be of great significance to estimate these 
uncertainties in order to complete our analysis. 
The largest uncertainty and probably the only one 
which can be relatively easy estimated comes 
from the errors on $\Delta Q$'s. As an example, we estimate 
the errors on $\Lambda$ production since we have had some available 
experimental data for making a comparison with our predictions. 
The uncertainties for other hyperon production are similar to 
the $\Lambda$ case. We have noticed that there still exist big 
errors on $\Delta Q$'s although their central values have been 
improved by adopting the new fit. In order to calculate the errors on 
$\Delta Q$'s, we express $\Delta U =\Delta D$ and $\Delta S$  
for the $\Lambda$ in terms of the five 
semileptonic hyperon decay constants ($A_i$, $i=1 \sim 5$) and $\Gamma_p$, 

\begin{equation}
\Delta U= \frac{11}{23}A_1 - \frac{37}{69} A_2+\frac{7}{46}A_3+
\frac{107}{46}A_4 +\frac{85}{23} A_5 -\frac{4}{23}\Gamma_p,
\end{equation}

\begin{equation}
\Delta S= -\frac{54}{23}A_1 - \frac{90}{23} A_2-\frac{243}{23}A_3-
\frac{639}{23}A_4 -\frac{66}{23} A_5 +\frac{51}{23}\Gamma_p.
\end{equation}
We estimate errors on $\Delta Q$'s by simply adding the all errors on the  
decay constants and $\Gamma_p$ in quadrature and we get 

\begin{equation}
\Delta U=\Delta D \simeq 0.03 \pm 0.19,
\end{equation}
and 

\begin{equation}
\Delta S \simeq 0.65 \pm 0.78.
\end{equation}
The errors on $\Delta Q$'s  are still large although their 
central values seem to be more reasonable than those in the original fit 
of Ref.~\cite{KimNPA}. With the errors, 
$\Delta S$ may vary in the range $[-0.13, 1.43]$ and $\Delta U =\Delta D$ 
in the range $[-0.16, 0.22]$. 
The value  $\Delta S= 0.62$ for the SU(3) symmetry case is in 
the  range of $\Delta S$ with errors for the SU(3) symmetry breaking case, 
but the value $\Delta U = \Delta D= -0.17$  for the SU(3) 
symmetry case  stays at the edge of the band of 
the $\Delta U$ ($\Delta D$) for the SU(3) symmetry breaking case. 
Therefore, the theoretical errors for our predictions are supposed 
to be very large.

Let us analyse the three processes for $\Lambda$ production before looking
into the effect of errors. First, in consideration of the fact the strange 
quark polarization is much larger than the $u$ and $d$ quark polarizations
in the $\Lambda$, and  according to the relative magnitudes of the 
constants $\hat{A}_q$ and $\hat{C}_q$ in Eq.~(\ref{PL2}) 
for $q=u$, $d$ and $s$, we have noticed that the strange quark 
dominates the $\Lambda$ polarization in $e^+e^-$ 
annihilation. Second, we find the spin transfer to $\Lambda$ 
in charged lepton DIS  can be approximated by

\begin{equation}
A_{\Lambda} \sim \frac {\Delta D_u^{\Lambda}(z)}{D_u^{\Lambda}(z)},
\end{equation}
due to the charge factor for the $u$ quark. Finally, the $\Lambda$ 
polarization in neutrino DIS is also mainly controlled by 

\begin{equation}
P_{\nu}^{\Lambda} \sim - \frac{ \Delta D_u^{\Lambda}(z)}{ D_u^{\Lambda}(z)}.
\end{equation}
To sum up, the $\Lambda$ polarization in $e^+e^-$ annihilation
is sensitive to the error on $\Delta S$'s, and the spin transfer
to $\Lambda$ in charged lepton DIS and the $\Lambda$ 
polarization in neutrino DIS are sensitive to the error on $\Delta U$'s.

Now, let us look the detailed numerical 
calculation results. With the allowed value ranges of 
$\Delta S$ and $\Delta U$ for the $\Lambda$, we estimate the errors 
for our predictions of the 
$\Lambda$ polarization in $e^+e^-$
annihilation, the spin transfer to $\Lambda$ in charged lepton DIS and 
the $\Lambda$ polarization in neutrino DIS.
The numerical results are  shown as bands in 
Figs.~\ref{a04f12}-\ref{a04f14}. The predictions with 
SU(3) symmetry are also shown in Figs.~\ref{a04f12}-\ref{a04f14} in dashed 
curves as a comparison. As expected, the theoretical errors 
for our predictions are indeed very large. 
However, the available experimental data seem to favor the predictions 
with SU(3) symmetry breaking even though the errors on $\Delta Q$'s 
are included. Therefore, the high precision measurements of the semileptonic 
hyperon decay constants and the hyperon polarizations are crucial important
in order to get a distinguishable evidence for SU(3) symmetry breaking 
from hyperon production.

\begin{acknowledgments}

This work is inspired by my cooperative
work with Bo-Qiang Ma, Ivan Schmidt, and Jacques Soffer. I would
like to express my great thanks to them for their encouragements
and valuable comments. In addition, this work is partially 
supported by National Natural Science Foundation of China 
under Grant Number 19875024 and by Fondecyt (Chile) project 3990048.

\end{acknowledgments}



\begin{thebibliography}{99}


\bibitem{KimPRD}
H. C. Kim, M. Praszalowicz, and K. Goeke, Phys. Rev. D {\bf 61},
114006 (2000).

\bibitem{Manohar98}
R. Flores-Mendieta, E. Jenkins, A. V. Manohar, Phys. Rev. D {\bf 58}, 94028 (1998).

\bibitem{KimNPA}
H. C. Kim, M. Praszalowicz, and K. Goeke, Nucl. Phys. A {\bf 691},
403 (2001); Acta Physica Polonica B {\bf 31}, 1767 (2000).

\bibitem{SigmaP}
Y.W.~Wah {\it et al}, Phys. Rev. Lett. {\bf 55}, 2551 (1985);
C.~Wilkinson {\it et al}, Phys. Rev. Lett. {\bf 58}, 855 (1987);
E761 Collaboration, A.~Morelos {\it et al.}, Phys. Rev. Lett. {\bf
71}, 2172 (1993).

\bibitem{XiP}
K.~Heller {\it et al.}, Phys. Rev. Lett. {\bf 51}, 2025 (1983);
P.M.~Ho {\it et al.}, Phys. Rev. Lett. {\bf 65}, 1713 (1990); J.
Duryea {\it et al.}, Phys. Rev. Lett. {\bf 67}, 1193 (1991).

\bibitem{Sigma0P}
B.E.~Bonner {\it et al.}, Phys. Rev. Lett. {\bf 62}, 1591 (1989).



\bibitem{GLR}
V.N.~Gribov and L.N.~Lipatov, Phys. Lett.  B {\bf 37}, 78 (1971);
Sov. J. Nucl. Phys.  {\bf 15}, 675 (1972); S.J.~Brodsky and
B.-Q.~Ma, Phys. Lett.  B {\bf 392}, 452 (1997).

\bibitem{DISprocess}
For a review, see, e.g., J.F. Owens and W.-K. Tung,
Ann. Rev. Nucl. Part. Sci. {\bf 42}, 291 (1992).

\bibitem{DYprocess}
For a review, see, e.g., P.L. McGaughey, J.M. Moss, and J.C. Peng,
Ann. Rev. Nucl. Part. Sci. {\bf 49}, 217 (1999).

\bibitem{Bhalerao00}
R. S. Bhalerao, N. G. Kelkar, B. Ram, Phys. Lett. B {\bf 476}, 285
(2000); R. S. Bhalerao, Phys. Rev. C {\bf 63}, 025208 (2001).

\bibitem{EMC89}
J. Ashman {\it et al.}, Nucl. Phys. B {\bf 328},1 (1989).

\bibitem{HSD1}
Particle Data Group, R. M. Barnett, C. D. Carone, D. E. Groom 
{\it et al.}, Phys. Rev. D {\bf 54}, 1 (1996).

\bibitem{HSD2}
M. Bourquin {\it et al.}, Z. Phys. C {\bf 12}, 307 (1982); {\bf
21},1 (1983).


\bibitem{Ma96}
B. Q. Ma, Phys. Lett. B {\bf 375}, 320 (1996).

\bibitem{MSSY}
B.-Q. Ma, I. Schmidt, and J.-J. Yang, Phys. Rev.  D {\bf 61},
034017 (2000); Phys. Lett.  B {\bf 477}, 107 (2000); B.-Q. Ma, I.
Schmidt, J. Soffer, and J.-J. Yang, Eur. Phys. J.  C {\bf 16}, 657
(2000); Phys. Lett.  B {\bf 488}, 254 (2000); Phys. Rev.  D {\bf
62}, 114009 (2000), Phys. Rev. D {\bf 64}, 014017 (2001); 
hep-ph/0107157. 

\bibitem{countingr}
R. Blankenbecler and S.J. Brodsky, Phys. Rev.  D {\bf 10}, 2973
(1974); J.F. Gunion, Phys. Rev.  D {\bf 10}, 242 (1974); S.J.
Brodsky and G.P. Lepage, in Proc. 1979 Summer Inst. on Particle
Physics, SLAC (1979).

\bibitem{Bro95}
S.J. Brodsky, M. Burkardt, and I. Schmidt, Nucl. Phys.  B {\bf
441}, 197 (1995).


\bibitem{Flambaum98}
V. V. Flambaum {\it{ et al.}}, Phys. Rev. E {\bf 57}, 4933 (1998).

\bibitem{Bickerstaff90}
R. P. Bickerstaff, J. T. Londergan, Phys. Rev. D {\bf 42}, 3621
(1990); F. Buccella {\it{ et al.}}, Mod. Phys. Lett. A {\bf 13},
441 (1998).

\bibitem{Bhalerao96}
R. S. Bhalerao, Phys. Lett. B {\bf 380}, 1 (1996); B {\bf 387},
881(E) (1996).


\bibitem{Yang99}
U.K. Yang and A. Bodek, Phys. Rev. Lett. {\bf 82}, 2467 (1999).


\bibitem{Sch00}
I.~Schmidt and J.-J.~Yang, hep-ph/0005054, Eur. Phys. J. C {\bf
20}, 63 (2001).


\bibitem{Bor98}
C.~Boros and Z.~Liang, Phys. Rev. D {\bf 57}, 4491 (1998).

\bibitem{Nza95}
R.~Jakob, P.J.~Mulders, J.~Rodrigues, Nucl. Phys.  A {\bf 626},
937 (1997); M.~Nzar and P.~Hoodbhoy, Phys. Rev. D {\bf 51}, 32
(1995).


\bibitem{Bor99}
C.~Boros and and A.W.~Thomas, Phys. Rev.  D {\bf 60}, 074017 (1999);
C.~Boros, T.~Londergan, and A.W.~Thomas, Phys. Rev. D {\bf  61}, 014007
(2000).

\bibitem{Yang02}
J. J. Yang, Phys. Rev. D {\bf 64}, 074010 (2001).



\bibitem{BRV00} J. Bl\"umlein, V. Radindran and W.L. van Neerven,
Nucl. Phys. B {\bf 589}, 349 (2000).

\bibitem{Bar00} V.~Barone, A.~Drago, and B.-Q.~Ma,
Phys. Rev. C {\bf 62}, 062201(R) (2000).

\bibitem{Yang01}
J.-J.~Yang, hep-ph/0107222, Phys. Lett. {\bf B 512}, 57 (2001).

\bibitem{Miyama94}
M. Miyama and S. Kumano, Comput. Phys. Commun. {\bf 94}, 185
(1996); M. Hirai, S. Kumano, and M. Miyama, {\it{ibid.}} {\bf
108}, 38 (1998).

\bibitem{ALEPH96}
ALEPH Collaboration, D. Buskulic {\it et al}, Phys. Lett.  B {\bf
374}, 319 (1996).

\bibitem{DELPHI95}
DELPHI Collaboration, Report No.DELPHI 95-86 PHYS 521,
CERN-PPE-95-172, presented at the EPS-HEP 95 conference, Brussels,
1995.

\bibitem{OPAL97}
OPAL Collaboration, K. Ackerstaff {\it et al},
Eur. Phys. J.  C {\bf 2}, 49 (1998).

\bibitem{HERMES}
HERMES Collaboration, A. Airapetian {\it et al.}, hep-ex/9911017,
Phys. Rev. D {\bf 64}, 112005 (2001).

\bibitem{E665}
E665 Collaboration, M. R. Adams {\it et al.}, Eur. Phys. J. C {\bf
17}, 263 (2000).

\bibitem{NOMAD}
NOMAD Collaboration: P.Astier {\it et al.},  Nucl. Phys. B {\bf
588}, 3 (2000), CERN-EP/2000-111; {\it{ibid.}} {\bf 605}, 3
(2001), CERN-EP/2001-028;

NOMAD Collaboration, C. Lachaud, Th\`ese de Doctorat Univ. Denis
Diderot, Paris, 3/5/2000; D. V. Naumov, hep-ph/0101325.


\bibitem{CTEQ5}
CTEQ Collaboration, H. L. Lai {\it{ et al.}}, Eur. Phys. J. C {\bf
12}, 375 (2000).

\bibitem{Ma99}
B.-Q. Ma and J. Soffer, Phys. Rev. Lett. {\bf 82}, 2250 (1999).


\nonfrenchspacing
\end{thebibliography}
\end{document}